\documentclass[journal,transmag]{IEEEtran}
\newcommand{\nutensor}{\ensuremath{\overline{\overline{\nu}}}}
\newcommand{\sigmatensor}{\ensuremath{\overline{\overline{\sigma}}}}

\renewcommand{\d}{\ensuremath{ \mathrm{d}} }

\newcommand{\A}{\ensuremath{\vec{A}}}

\newcommand{\AvecComplex}{\ensuremath{\underline{\vec{A}}}}

\newcommand{\Js}{\ensuremath{\vec{J}_{\mathrm{s}}}}

\newcommand{\JsvecComplex}{\ensuremath{\underline{\vec{J}}_{\mathrm{s}}}}

\renewcommand{\epsilon}{\varepsilon}

\newcommand{\Omegals}{\Omega_{\text{ls}}} 
\newcommand{\OmegaC}{\Omega_{\text{c}}}
\newcommand{\OmegaS}{\Omega_{\text{s}}}

\newcommand{\dV}{\ensuremath{ \mathrm{d}V}}

\newcommand{\ff}{\ensuremath{f_{\mathrm{f}}}} 
\newcommand{\wf}{\ensuremath{\omega_{\mathrm{f}}}} 

\newcommand{\adjustedaccent}[1]{%
	\mathchoice{}{}
	{\mbox{\raisebox{-.5ex}[0pt][0pt]{$\scriptstyle#1$}}}
	{\mbox{\raisebox{-.35ex}[0pt][0pt]{$\scriptscriptstyle#1$}}}
}

\newcommand{\adjustedaccentsecondbow}[1]{%
	\mathchoice{}{}
	{\mbox{\raisebox{-.7ex}[0pt][0pt]{$\scriptstyle#1$}}}
	{\mbox{\raisebox{-.35ex}[0pt][0pt]{$\scriptscriptstyle#1$}}}
}

\newcommand\bow[1]{\overset{\adjustedaccent{\smallfrown}}{#1}}
\newcommand\secondbow[1]{\overset{\adjustedaccentsecondbow{\smallfrown}}{#1}}

\newcommand{\ahatc}[1]{\ensuremath{\bow{\underline{\mathbf{a}}}_{#1}}} 

\newcommand*\circled[1]{\tikz[baseline=(char.base)]{
		\node[shape=circle,draw,inner sep=2pt] (char) {#1};}}
        
\newcommand{\rv}{\ensuremath{\vec{r}}}

\newcommand{\deltaH}{\ensuremath{\delta_{\mathrm{H}}}}
\newcommand{\kH}{\ensuremath{k_{\mathrm{H}}}}
\newcommand{\deltaB}{\ensuremath{\delta_{\mathrm{B}}}}
\newcommand{\kB}{\ensuremath{k_{\mathrm{B}}}}
\usepackage{amssymb}
\usepackage{amsmath}
\usepackage{bm}
\usepackage{mathtools}
\usepackage{tikz}
\usetikzlibrary{patterns}
\usepackage{pgfplots}
\pgfplotsset{compat=1.18}
\usepackage{siunitx}
\usepackage{booktabs}
\usepackage{multirow}
\usepackage{standalone}
\usepackage{float}

\ifCLASSOPTIONcompsoc
\usepackage[caption=false,font=normalsize,labelfon
t=sf,textfont=sf]{subfig}
\else
\usepackage[caption=false,font=footnotesize]{subfig}
\fi

\usepackage{nicefrac}
\usepackage{xspace}
\usepackage{graphicx}
\usepackage{textcomp} 
\usepackage{latexsym} 
\usepackage{comment}
\usetikzlibrary{shapes.geometric, arrows, positioning,shapes.multipart}
\definecolor{verylightblue}{RGB}{0,156,218}
\definecolor{verylightpink}{RGB}{201,48,142}
\definecolor{red}{RGB}{230,0,26}
\definecolor{green}{RGB}{153,192,0}
\definecolor{blue}{RGB}{0,90,169}
\definecolor{lightblue}{RGB}{0,131,204}
\definecolor{lightsmaragd}{RGB}{0,131,204}
\definecolor{orange}{RGB}{236,101,0}
\definecolor{lightpurple}{RGB}{166,0,132}

\definecolor{darkblue_one}{RGB}{0,78,138} 
\definecolor{darkblue_two}{RGB}{0,104,157} 

\definecolor{darkemerald}{RGB}{0,136,119} 
\definecolor{darkgreen_one}{RGB}{127,171,22} 
\definecolor{darkgreen}{RGB}{127,171,22} 

\definecolor{darkgreen_two}{RGB}{177,189,0} 
\definecolor{darkyellow_one}{RGB}{215,172,0} 
\definecolor{darkyellow_two}{RGB}{210,135,0} 
\definecolor{darkyellow}{RGB}{210,135,0} 

\definecolor{darkorange}{RGB}{204,76,3} 

\definecolor{darkred}{RGB}{185,15,34} 

\definecolor{darkpurple}{RGB}{149,17,105} 

\tikzstyle{start} = [rectangle, rounded corners, minimum width=4.2cm, minimum height=0.5cm, text centered, draw=darkred, line width = 0.6mm]
\tikzstyle{stop} = [rectangle, rounded corners, minimum width=4.2cm, minimum height=0.5cm, text centered, draw=darkblue_two, line width = 0.6mm]
\tikzstyle{process_blue} = [rectangle, minimum width=4.2cm, minimum height=1cm, text centered, draw=darkblue_two, line width = 0.6mm]
\tikzstyle{process_red} = [rectangle, minimum width=4.2cm, minimum height=1cm, text centered, draw=darkred, line width = 0.6mm]
\tikzstyle{process_black} = [rectangle, minimum width=4.2cm, minimum height=1cm, text centered, draw=black, line width = 0.6mm]
\tikzstyle{decision} = [diamond, aspect=4, minimum width=3cm, minimum height=0.1cm, text centered, draw=darkblue_two, line width = 0.6mm]
\tikzstyle{arrow} = [thick,->,>=stealth]

\begin{document}

\title{DC-Biased Homogenized Harmonic Balance Finite Element Method}

\author{\IEEEauthorblockN{Jan-Magnus Christmann\IEEEauthorrefmark{1},
Laura A. M. D'Angelo\IEEEauthorrefmark{1},
Herbert De Gersem\IEEEauthorrefmark{1}, Sven Pfeiffer\IEEEauthorrefmark{2}, and Sajjad H. Mirza\IEEEauthorrefmark{2}}

\IEEEauthorblockA{\IEEEauthorrefmark{1}Institute for Accelerator Science and Electromagnetic Fields, Technical University of Darmstadt, Germany}
\IEEEauthorblockA{\IEEEauthorrefmark{2}Deutsches Elektronen-Synchrotron DESY, Hamburg, Germany}
\thanks{Corresponding author: J.-M. Christmann (email: jan-magnus.christmann@tu-darmstadt.de).
	This work was supported by Deutsches Elektronen-Synchrotron DESY, Hamburg, Germany.}
}

\markboth{Journal of \LaTeX\ Class Files,~Vol.~14, No.~8, August~2015}%
{Shell \MakeLowercase{\textit{et al.}}: Bare Demo of IEEEtran.cls for IEEE Transactions on Magnetics Journals}
%



\IEEEtitleabstractindextext{
	\vspace{-2em}
	\begin{center}
		\small\itshape
		This work has been submitted to the IEEE for possible publication. 
		Copyright may be transferred without notice, after which this version may no longer be accessible.
	\end{center}
	\vspace{1em}
\begin{abstract}
The homogenized harmonic balance finite element (FE) method enables efficient nonlinear eddy-current simulations of 3-D devices with lamination stacks by combining the harmonic balance method with a frequency-domain-based homogenization technique. This approach avoids expensive time stepping of the eddy-current field problem and allows the use of a relatively coarse FE mesh that does not resolve the individual laminates. In this paper, we extend the method to handle excitation signals with a dc bias. To achieve this, we adapt the original homogenization technique to better account for ferromagnetic saturation. The resulting formula for the homogenized reluctivity is evaluated using a look-up table computed from a 1-D FE simulation of a lamination and containing the average magnetic flux density in the lamination and the corresponding skin depth. We compare the results of the proposed method to those from a fine-mesh transient reference simulation. The tests cover different levels of ferromagnetic saturation and frequencies between 50\,Hz and 10\,kHz.
For moderate ferromagnetic saturation, the method gives a good approximation of the eddy-current losses and the magnetic energy, with relative errors below 10\%, while reducing the required number of degrees of freedom at 10\,kHz by 1.5 orders of magnitude. This results in a reduction in simulation time from 2 days on a contemporary server to 90 minutes on a standard workstation.
\end{abstract}

}

\maketitle

\IEEEdisplaynontitleabstractindextext

%
\IEEEpeerreviewmaketitle

\section{Introduction}\label{sec:intro}
\IEEEPARstart{W}{hen} simulating 3-D inductor models with laminated yokes or cores, the finite-element (FE) mesh does not only have to resolve the individual laminations, which usually have a thickness of $\SI{1}{\milli\meter}$ or less, but also the skin depth $\delta$, which scales with the frequency $f$ according to $\nicefrac{1}{\sqrt{f}}$. Hence, for simulations at frequencies in the kilohertz range, the number of degrees of freedom (DoFs) necessary in a brute-force approach becomes prohibitive for most realistic models and one must apply a homogenization technique~\cite{sabariego_2020}.

If a nonlinear $B$--$H$ curve is considered, the only available option is to resort to multiscale methods. There are different variants of such methods, but the main idea is always to decompose the problem into a macroscale problem, discretized on a coarse FE mesh, and many microscale problems discretized on fine FE meshes~\cite{e_2003,efendiev_2004}. The macroscale problem and the microscale problems are coupled and are solved iteratively, which typically results in rather complex numerical schemes that are not well-suited for large 3-D models.

To address this issue, we have developed the homogenized harmonic balance finite element method (HomHBFEM)~\cite{christmann_2025aa}, which combines a frequency-domain-based homogenization~\cite{dular_2003aa} for the laminations with the harmonic balance finite element method (HBFEM)~\cite{yamada_1989}. In this way, ferromagnetic saturation is included without costly time-stepping, and the eddy-current effects in the laminations are captured without the need for a very large FE mesh. 

\begin{figure}[t]
    \centering
    \includegraphics[width=\linewidth]{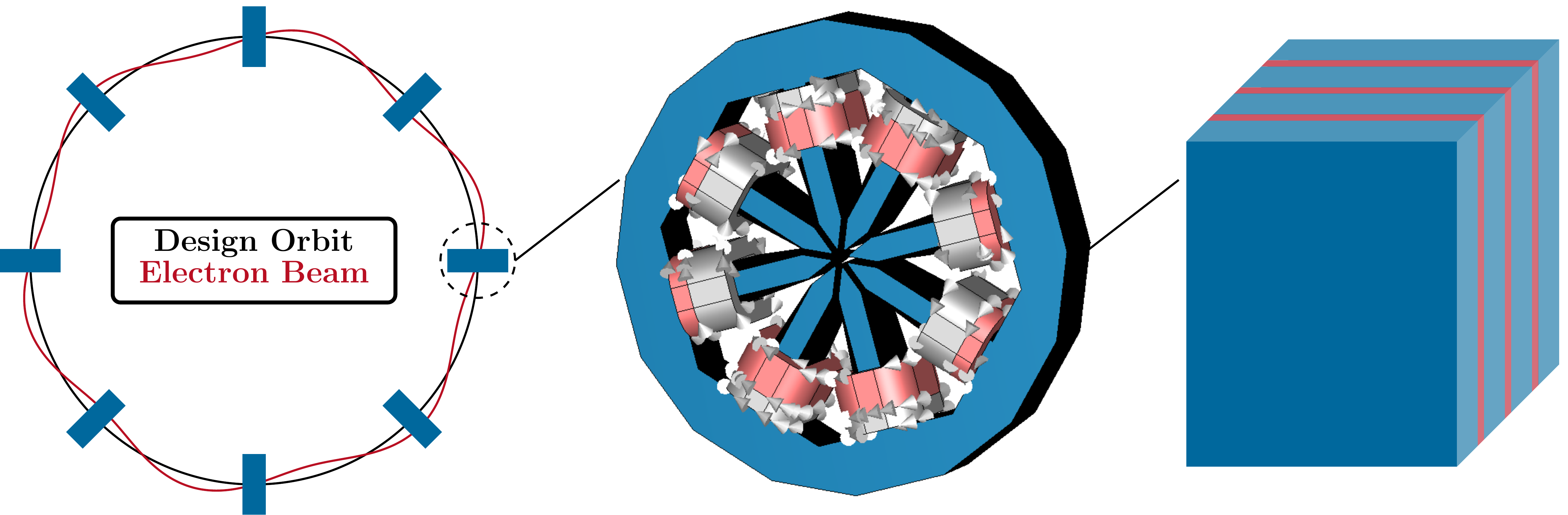}
    \caption{Electron beam in a particle accelerator (left) with FC magnets (center) and their laminations (right).}
    \label{fig:fc_magnet}
\end{figure}

The HomHBFEM has enabled, for the first time, nonlinear eddy-current simulations of the fast corrector (FC) magnets (Fig.~\ref{fig:fc_magnet}) for \mbox{PETRA IV}, the next-generation synchrotron radiation source at Deutsches Elektronen-Synchrotron DESY~\cite{schroer_2019}. Hundreds of these FC magnets will be distributed around the $\SI{2.3}{\kilo\meter}$ circumference of the particle accelerator to correct deviations of the electron beam from the design orbit. They will be part of a feedback system and combine fast and slow correction in one magnet. Hence, the excitation current will have dc- and ac-components up to the kilohertz range, which are determined during operation by a control algorithm based on measurements of the deviations from the design orbit~\cite{mirza_2023}. So far, the HomHBFEM was only applicable in case of a pure ac excitation. In this work, we extend the range of applicability to a harmonic excitation with a dc bias in order to enable an investigation of the FC magnets' dynamic behavior with a more realistic excitation.

The remainder of this article is structured as follows. In Section~\ref{sec:method_standard}, we briefly introduce the nonlinear \mbox{magnetoquasistatic (MQS)} problem and the HBFEM, before explaining the standard HomHBFEM. Section~\ref{sec:method_dc_biased} deals with the extension of the HomHBFEM to the dc-biased case. Although a naive extension is possible and theoretically straightforward, it performs poorly in practice. Therefore, we propose a more refined approach, which forms the main contribution of this paper. In Section~\ref{sec:verification}, the refined approach is verified by comparing its results for a 3-D laminated inductor model to a transient reference simulation conducted in CST Studio Suite\textsuperscript{\textregistered} using an adequately fine FE mesh. In this context, we also show the superior performance of the refined approach compared to the naive one. Finally, Section~\ref{sec:conclusion} concludes this article.

\section{Standard HomHBFEM}\label{sec:method_standard}
\subsection{HBFEM}
The strong form of the $\A^{*}$-formulation~\cite{emson_1983} of the nonlinear MQS problem reads
\begin{equation}\label{eq:MQS_NL_TD}
	\nabla \times \left(\nu\left( \rv, t \right)  \nabla\times \vec{A}\left(\rv,t\right) \right) + \sigma(\rv) \frac{\partial \vec{A} \left(\rv, t \right)}{\partial t} = \vec{J}_{\mathrm{s}}\left(\rv, t\right),
\end{equation}
where $t$ indicates the time, $\rv$ the position vector, $\sigma$ the conductivity, $\nu$ the reluctivity, $\Js$ the source current density, and the unknown $\A$ is the magnetic vector potential. The magnetic flux density $\vec{B}$ can be computed as $\vec{B} = \nabla \times \vec{A}$ and is related to the magnetic field strength $\vec{H}$ via $\vec{H} = \nu \vec{B}$. The reluctivity depends on time since it depends on the solution due to the nonlinearity of the $B$--$H$ curve. Note that we only consider non-hysteretic $B$--$H$ curves throughout this work.

To numerically approximate the solution of Eq.~\eqref{eq:MQS_NL_TD}, the corresponding weak formulation is discretized in space with edge elements according to the FE method~\cite{biro_1999}.

To avoid costly time-stepping of the discretized problem, the HBFEM transforms the problem into frequency domain and uses a multiharmonic approach to account for the nonlinearity. This approach approximates the solution with a Fourier series truncated at a certain maximum harmonic order $m \in \mathbb{N}$, leading to the complex-valued system of equations
\begin{equation}\label{eq:HBFEM}
	\sum_{k = \max\{-m,n-m\} }^{\min\{m, n+m\}} \mathbf{K}_{\nu_k} \left( \bow{\underline{\mathbf{a}}}\right)\bow{\underline{\mathbf{a}}}_{n-k} + \jmath n\wf\mathbf{M}_{\sigma}\bow{\underline{\mathbf{a}}}_{n} = \secondbow{\bow{\underline{\mathbf{j}}}}_{\mathrm{s},n},
\end{equation}
for $ n \in \mathbb{Z} \, \cap \, [-m,m]$~\cite{roppert_2019}. In this equation, $\wf$ is the fundamental angular frequency, $\bow{\underline{\mathbf{a}}}_{n}$ is the vector gathering the DoFs of the $n$-th harmonic of the magnetic vector potential, $\secondbow{\bow{\underline{\mathbf{j}}}}_{\mathrm{s},n}$ is the discretized $n$-th harmonic of the source current density, $\mathbf{M}_{\sigma}$ is the mass matrix, and  $\mathbf{K}_{\nu_k}$ denotes the stiffness matrix computed with the $k$-th harmonic of the reluctivity. We indicate that each harmonic of the reluctivity depends on all the harmonics of the magnetic vector potential by using the notation $\mathbf{K}_{\nu_k}\left(\bow{\underline{\mathbf{a}}}\right)$.

Considering that $\bow{\underline{\mathbf{a}}}_{-n} = \bow{\underline{\mathbf{a}}}_{n}^{*}$, one only has to solve for $\left( \bow{\underline{\mathbf{a}}}_{n} \right)_{n \in \mathbb{N}_0}$. Moreover, if $\Js$ only contains odd harmonics of the fundamental component, $\A$ will also only include odd harmonics~\cite{bachinger_2005}. In this case, the system of equations~\eqref{eq:HBFEM} is further reduced in size~\cite{degersem_2001}. For a more detailed explanation and a step-by-step derivation of Eq.~\eqref{eq:HBFEM}, we refer the reader to~\cite{christmann_2025aa}.

\subsection{HomHBFEM Without DC Bias}
Choosing the coordinate system as in Fig.~\ref{fig:homogenization}, the homogenization~\cite{dular_2003aa} transforms the laminated yokes or cores into bulk models by replacing $\sigma(\vec{r})$ and $\nu(\vec{r})$ with spatially constant material tensors
\begin{align}
	\sigmatensor &= \sigma \begin{bmatrix}
		1 & 0 & 0 \\ 0 & 1 & 0 \\ 0 & 0 & 0
	\end{bmatrix}, \label{eq:tensor_conductivity} \\
	\begin{split}
		\underline{\nutensor} &=  \frac{\sigma d \delta \omega \left(1+\jmath\right)}{8} \frac{\sinh \left( \left(1 + \jmath \right) \frac{d}{\delta}\right)}{\sinh^2\left( \left(1 + \jmath \right) \frac{d}{2\delta}\right)} 
		\begin{bmatrix}
			1 & 0 & 0 \\ 0 & 1 & 0 \\ 0 & 0 & 0 
		\end{bmatrix} \\  & + \nu \begin{bmatrix}
			0 & 0 & 0 \\ 0 & 0 & 0 \\ 0 & 0 & 1
		\end{bmatrix}. \label{eq:tensor_reluctivity}
	\end{split}
\end{align}
Herein, $d$ is the thickness of the conductive laminates, $\nu$ is their reluctivity, $\sigma$ their conductivity and
$\delta = \sqrt{\frac{2\nu}{\sigma\omega}}$ is the skin depth.

Equations~\eqref{eq:tensor_conductivity} and~\eqref{eq:tensor_reluctivity} asssume that the insulation thickness is negligible compared to the lamination thickness and that the magnetic flux perpendicular to the lamination is insignificant. However, the method can be easily adapted for the case of non-negligible insulation thickness and perpendicular magnetic flux according to the following well-established standard approach, see e.g.~\cite{silva_1995, degersem_2012,kaimori_2007}.

We replace $\sigma$ with $\tilde{\sigma} = \gamma \sigma$, where $\gamma$ is the stacking factor, meaning the percentage of the laminated volume consisting of conductive material. In the $x$- and $y$-components of the reluctivity tensor in Eq.~\eqref{eq:tensor_reluctivity}, we replace $\nu$ within the skin depth formula with 
\begin{equation}\label{eq:nu_xy}
	\tilde{\nu}_{xy} = \frac{1}{\left( 1 - \gamma \right) \nu_{\mathrm{ins}}^{-1} + \gamma \nu^{-1}}, 
\end{equation}
where $\nu_{\mathrm{ins}}$ is the reluctivity of the insulation. 
For the $z$-component of the reluctivity tensor, we use 
\begin{equation}\label{eq:nu_z}
    \tilde{\nu}_z = \gamma \nu + \left(1 - \gamma\right)\nu_{\mathrm{ins}}.
\end{equation}
\begin{figure}[t]
	\centering
	\resizebox{0.355\textwidth}{!}{%
			
	\begin{tikzpicture}
		\def\width{5}        
		\def\height{6}       
		\def\blueDepth{1}    
		\def\orangeDepth{0.3} 
		
		\def\zcoord{0}
		
		\fill[darkblue_two!100] (0, 0, \zcoord) -- ++(\width, 0, 0) -- ++(0, \height, 0) -- ++(-\width, 0, 0) -- cycle;
		\fill[darkblue_two!70] (0, \height, \zcoord) -- ++(\width, 0, 0) -- ++(0, 0, \blueDepth) -- ++(-\width, 0, 0) -- cycle;
		\fill[darkblue_two!60] (\width, 0, \zcoord) -- ++(0, \height, 0) -- ++(0, 0, \blueDepth) -- ++(0, -\height, 0) -- cycle;
		\pgfmathsetmacro{\zcoord}{\zcoord + \blueDepth} 
		
		\fill[darkred!100] (0, 0, \zcoord) -- ++(\width, 0, 0) -- ++(0, \height, 0) -- ++(-\width, 0, 0) -- cycle;
		\fill[darkred!70] (0, \height, \zcoord) -- ++(\width, 0, 0) -- ++(0, 0, \orangeDepth) -- ++(-\width, 0, 0) -- cycle;
		\fill[darkred!60] (\width, 0, \zcoord) -- ++(0, \height, 0) -- ++(0, 0, \orangeDepth) -- ++(0, -\height, 0) -- cycle;
		\pgfmathsetmacro{\zcoord}{\zcoord + \orangeDepth} 
		
		\fill[darkblue_two!100] (0, 0, \zcoord) -- ++(\width, 0, 0) -- ++(0, \height, 0) -- ++(-\width, 0, 0) -- cycle;
		\fill[darkblue_two!70] (0, \height, \zcoord) -- ++(\width, 0, 0) -- ++(0, 0, \blueDepth) -- ++(-\width, 0, 0) -- cycle;
		\fill[darkblue_two!60] (\width, 0, \zcoord) -- ++(0, \height, 0) -- ++(0, 0, \blueDepth) -- ++(0, -\height, 0) -- cycle;
		\pgfmathsetmacro{\zcoord}{\zcoord + \blueDepth} 
		
		\fill[darkred!100] (0, 0, \zcoord) -- ++(\width, 0, 0) -- ++(0, \height, 0) -- ++(-\width, 0, 0) -- cycle;
		\fill[darkred!70] (0, \height, \zcoord) -- ++(\width, 0, 0) -- ++(0, 0, \orangeDepth) -- ++(-\width, 0, 0) -- cycle;
		\fill[darkred!60] (\width, 0, \zcoord) -- ++(0, \height, 0) -- ++(0, 0, \orangeDepth) -- ++(0, -\height, 0) -- cycle;
		\pgfmathsetmacro{\zcoord}{\zcoord + \orangeDepth} 
		
		\fill[darkblue_two!100] (0, 0, \zcoord) -- ++(\width, 0, 0) -- ++(0, \height, 0) -- ++(-\width, 0, 0) -- cycle;
		\fill[darkblue_two!70] (0, \height, \zcoord) -- ++(\width, 0, 0) -- ++(0, 0, \blueDepth) -- ++(-\width, 0, 0) -- cycle;
		\fill[darkblue_two!60] (\width, 0, \zcoord) -- ++(0, \height, 0) -- ++(0, 0, \blueDepth) -- ++(0, -\height, 0) -- cycle;
		\pgfmathsetmacro{\zcoord}{\zcoord + \blueDepth} 
		
		\fill[darkred!100] (0, 0, \zcoord) -- ++(\width, 0, 0) -- ++(0, \height, 0) -- ++(-\width, 0, 0) -- cycle;
		\fill[darkred!70] (0, \height, \zcoord) -- ++(\width, 0, 0) -- ++(0, 0, \orangeDepth) -- ++(-\width, 0, 0) -- cycle;
		\fill[darkred!60] (\width, 0, \zcoord) -- ++(0, \height, 0) -- ++(0, 0, \orangeDepth) -- ++(0, -\height, 0) -- cycle;
		\pgfmathsetmacro{\zcoord}{\zcoord + \orangeDepth} 
		
		\fill[darkblue_two!100] (0, 0, \zcoord) -- ++(\width, 0, 0) -- ++(0, \height, 0) -- ++(-\width, 0, 0) -- cycle;
		\fill[darkblue_two!70] (0, \height, \zcoord) -- ++(\width, 0, 0) -- ++(0, 0, \blueDepth) -- ++(-\width, 0, 0) -- cycle;
		\fill[darkblue_two!60] (\width, 0, \zcoord) -- ++(0, \height, 0) -- ++(0, 0, \blueDepth) -- ++(0, -\height, 0) -- cycle;
		
		\pgfmathsetmacro{\zcoord}{\zcoord + \blueDepth} 
		\fill[darkblue_two!100] (0, 0, \zcoord) -- ++(\width, 0, 0) -- ++(0, \height, 0) -- ++(-\width, 0, 0) -- cycle;
		
		\draw[->, line width= 2pt] (-3.1, -2) -- (-3.1, -0.8);
		\node at (-3.1, -0.8) [anchor=east] {\huge $y$};
		
		\draw[->, line width= 2pt] (-3.1, -2) -- (-1.9, -2);
		\node at (-1.9, -2.1) [anchor=north] {\huge $x$};

		\draw[line width= 2pt] (-3.1,-2) circle [radius=0.2cm];
		\filldraw (-3.1,-2) circle (2.5pt);
		\node at (-3.2, -2.35) [anchor=east] {\huge $z$};
		
		\draw[->,line width = 2pt] (-0.5,6.5)--(0,5.8);
		\node at (-1.5,6.8) [anchor=west] {\huge $\nu_{\mathrm{c}}, \sigma_{\mathrm{c}}  $};
		
		\draw[->, line width = 2pt] (-1.3,5.75)--(-0.7,5.05);
		\node at (-2.9,5.95) [anchor=west] {\huge $\nu_{\mathrm{ins}}$, $\sigma_{\mathrm{ins}}$};

		\def\zcoord{0}
		\def\xcoord{9}
		\def\Depth{4.9}    
		\fill[darkblue_two!100] (\xcoord, 0, \zcoord) -- ++(\width, 0, 0) -- ++(0, \height, 0) -- ++(-\width, 0, 0) -- cycle;
		\fill[darkblue_two!70] (\xcoord, \height, \zcoord) -- ++(\width, 0, 0) -- ++(0, 0, \Depth) -- ++(-\width, 0, 0) -- cycle;
		\fill[darkblue_two!60] (\xcoord + \width, 0, \zcoord) -- ++(0, \height, 0) -- ++(0, 0, \Depth) -- ++(0, -\height, 0) -- cycle;

		\pgfmathsetmacro{\zcoord}{\zcoord + \Depth} 
		\fill[darkblue_two!100] (\xcoord, 0, \zcoord) -- ++(\width, 0, 0) -- ++(0, \height, 0) -- ++(-\width, 0, 0) -- cycle;
		
		
		\draw[->, line width = 2pt] (5+ \xcoord,6.5)--(4.5 + \xcoord,5.8);
		\node at (4.5+ \xcoord,7.0) [anchor=west] {\huge $\underline{\overline{\overline{\nu}}}, \overline{\overline{\sigma}}$};
		
		
		\draw[->, line width = 2pt] (\xcoord -5, 5.8) arc [start angle = 180,end angle =0, x radius = 2.9, y radius =1]; 
		\node at (\xcoord - 2,7) [anchor = south] {\huge homogenization technique};
		
	\end{tikzpicture}
	%
	}
	\caption{Left: Lamination stack with insulation in red and conducting laminates in blue. Right: Homogenized model.}
	\label{fig:homogenization}
\end{figure}
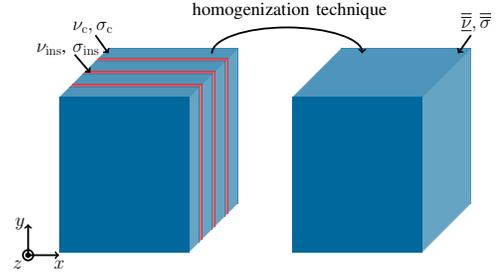Finally, $d$ is replaced with $\tilde{d} = d + d_{\mathrm{ins}}$ where $d_{\mathrm{ins}}$ is the thickness of the insulation.

To understand how the HBFEM is combined with the homogenization technique, we rewrite Eq.~\eqref{eq:HBFEM} as
\begin{align}
	\mathbf{K}_{\nu_0}\left(\bow{\underline{\mathbf{a}}}\right)\bow{\underline{\mathbf{a}}}_{n} &+ \sum_{\substack{k= \max\{-m,n-m\}, \\ k \neq 0}}^{\min\{m, n+m\}} \mathbf{K}_{\nu_k} \left( \bow{\underline{\mathbf{a}}}\right)\bow{\underline{\mathbf{a}}}_{n-k} \notag \\  &+ \jmath n\wf\mathbf{M}_{\sigma}\bow{\underline{\mathbf{a}}}_{n} = \secondbow{\bow{\underline{\mathbf{j}}}}_{\mathrm{s},n}.
\end{align}
Now, the material tensors of the homogenization technique are introduced. The introduction of the conductivity tensor $\sigmatensor$ into the mass matrix $\mathbf{M}_{\sigma}$ is straightforward. The reluctivity tensor $\underline{\nutensor}$, on the other hand, is only used in the construction of $\mathbf{K}_{\nu_0}$, i.e., only for the stiffness matrix containing the dc component of the reluctivity. Thereby, we obtain
\begin{align}\label{eq:homhbfem}
	\mathbf{K}_{\nutensor_0 \left(n\wf \right)}\left(\bow{\underline{\mathbf{a}}}\right)\bow{\underline{\mathbf{a}}}_{n} &+ \sum_{\substack{k= \max\{-m,n-m\}, \\ k \neq 0}}^{\min\{m, n+m\}} \mathbf{K}_{\nu_k} \left( \bow{\underline{\mathbf{a}}}\right)\bow{\underline{\mathbf{a}}}_{n-k} \notag \\ &+ \jmath n\wf\mathbf{M}_{\sigmatensor}\bow{\underline{\mathbf{a}}}_{n} = \secondbow{\bow{\underline{\mathbf{j}}}}_{\mathrm{s},n},
\end{align}
where $\mathbf{M}_{\sigmatensor}$ is the mass matrix constructed with the conductivity tensor $\sigmatensor$ as given in Eq.~\eqref{eq:tensor_conductivity} and $\mathbf{K}_{\nutensor_0 \left(n\wf \right)}$ is the stiffness matrix constructed with the reluctivity tensor given in Eq.~\eqref{eq:tensor_reluctivity} evaluated at $w = n \wf$ and $\nu = \nu_0$ for each element.

The system \eqref{eq:homhbfem} is linearized using successive substitution. To avoid solving multiharmonic systems of equations increasing in size according to the number of considered harmonics, we adopt a block Jacobi iteration~\cite{saad_2003}, leading to 
\begin{align}\label{eq:homhbfem_iteration}
	& \left( \mathbf{K}_{\nutensor_0 \left(n\wf \right)}(\bow{\underline{\mathbf{a}}}^{i}) + \jmath n\wf\mathbf{M}_{\sigmatensor}   \right)    \bow{\underline{\mathbf{a}}}^{i + 1}_{n} \notag\\ &= \secondbow{\bow{\underline{\mathbf{j}}}}_{\mathrm{s},n} - \sum_{\substack{k= \max\{-m,n-m\}, \\ k \neq 0}}^{\min\{m, n+m\}} \mathbf{K}_{\nu_k} ( \bow{\underline{\mathbf{a}}}^{i})\bow{\underline{\mathbf{a}}}^{i}_{n-k}, 
\end{align}
where the superscript $i = 0,1,2,...$ indicates the iteration number. In this way, the method is easily parallelized, i.e., the equations for the different harmonics can be solved in parallel in each iteration~\cite{yamada_1991}.

To compute the stiffness matrices in each iteration, we first need to compute the discretized magnetic flux densities $\secondbow{\bow{\underline{\mathbf{b}}}}_{n}^{\raisebox{-5pt}{\scriptsize $i$}}$ as the discrete curl of the $\bow{\underline{\mathbf{a}}}_{n}^{i}$. These complex-valued magnetic flux densities are transformed into the time domain. The time signal for the magnitude of the magnetic flux density is then inserted into the nonlinear $B$-$H$ curve to obtain the time signal for the magnitude of the magnetic field strength. By forming the quotient of these two time signals, we obtain the reluctivity~$\nu^{i}(t)$. Finally, computing the fast Fourier transform (FFT) of $\nu^{i}(t)$ gives us the Fourier series coefficients of the reluctivity, which allows us to set up $\mathbf{K}_{\nutensor_0 \left(n\wf \right)}(\bow{\underline{\mathbf{a}}}^{i})$ and $\mathbf{K}_{\underline{\nu}_k}(\bow{\underline{\mathbf{a}}}^{i})$ and to assemble the system of equations in \eqref{eq:homhbfem_iteration}.

This iteration will give us magnetic flux densities parallel to the lamination that are averaged over the lamination thickness.  We transform these averaged parallel components, $\underline{B}_{\parallel}^{\mathrm{a}}(z)$, back to local ones via
\begin{equation}\label{eq:trafo}
	\underline{B}_{\parallel}(z) =  \frac{\underline{B}_{\parallel}^{\mathrm{a}} (z) k d }{2 \sinh\left(k \frac{d}{2} \right)} \cosh\left( k \Tilde{z}\right),
\end{equation}
 with $k = \frac{1 + \jmath}{\delta}$, and $\Tilde{z}$ being the local coordinate within one lamination, where the origin of the local coordinate system is chosen in the center of the lamination. We evaluate the transformation in Eq.~\eqref{eq:trafo} in the $i$-th iteration by inserting the dc component of the reluctivity $\nu_0^{i}$ and the parallel magnetic flux density components $\underline{B}_{\parallel}^{\mathrm{a},i}(z)$.
\begin{figure}[t]
	\centering
	\begin{tikzpicture}[node distance=1.5cm, scale=0.85, every node/.style={transform shape}]]
		\node (start) [start] {Initialize $\ahatc{n}^{0}, \secondbow{\bow{\underline{\mathbf{b}}}}_{n}^{\raisebox{-5pt}{$\scriptstyle 0$}}$};
		\node (pro1) [process_blue, below=1.8cm of start, align=center] {Transform  $\secondbow{\bow{\underline{\mathbf{b}}}}_{n}^{\raisebox{-5pt}{$\scriptstyle i$}}$ into\\ time domain};
		\node (pro2) [process_blue, below of=pro1, align=center] {Compute $\nu^{i}\left(t\right)$ with\\nonlinear $B$-$H$ curve};
		
		\node (pro3) [process_blue, below=0.4cm of pro2, align=center] {FFT of $\nu^{i}\left( t\right)$};
		\node (pro4) [process_blue, right= 1.2cm of pro3, align=center] {Compute reluctivity \\ tensors $\nutensor_{0}^{i}\left(n\wf\right)$};
		\node (pro5) [process_red, above=0.4cm of pro4, align=center] {Assemble system \eqref{eq:homhbfem_iteration} \\  \& solve  $\rightarrow \ahatc{n}^{i+1}, \secondbow{\bow{\underline{\mathbf{b}}}}_{n}^{\raisebox{-5pt}{$\scriptstyle i + 1$}}$};
        \node (pro6) [process_blue, above = 0.4 cm of pro5, align = center]{Transformation formula~\eqref{eq:trafo}};
		\node (dec1) [decision, above=0.4cm of pro6, node distance=1.5cm] {Convergence?};
		\node (stop) [stop, above=0.4cm of dec1, node distance=1.5cm] {Stop};
		
		\draw [arrow] (start) -- (pro1);
		\draw [arrow] (pro1) -- (pro2);
		\draw [arrow] (pro2) -- (pro3);
		\draw [arrow] (pro3) -- (pro4);
		\draw [arrow] (pro4) -- (pro5);
		\draw [arrow] (pro5) -- (pro6);
		\draw [arrow] (pro6) -- (dec1);
        \draw [arrow] (dec1) -- (stop);
        
		\node[anchor=west] at (5.7,-0.5) {yes};
		\draw (3.23,-1.23) -- (0, -1.23) ;
	
		\node[anchor=north] at (1.5, -1.3) {$\small i = i + 1$};
        \node[anchor=south] at (1.5, -1.23) {no};
	\end{tikzpicture}
	\caption{Flow chart of the HomHBFEM. The index $n$ indicates the harmonic order and superscript $i$ the iteration number. Blue boxes are implemented in Python, red boxes in GetDP.}\label{fig:flowchart}
\end{figure}
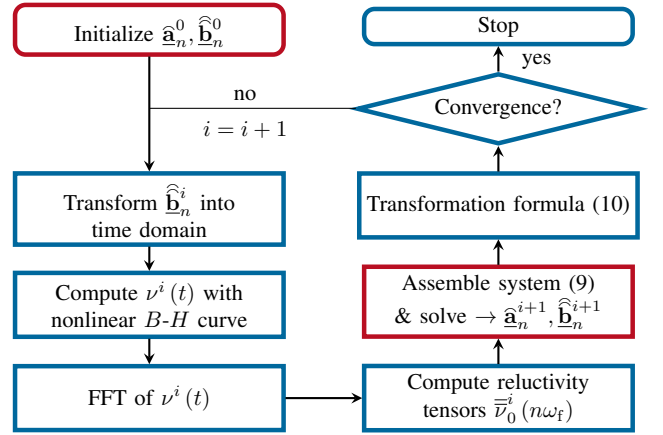

The iteration is stopped upon fulfillment of an energy-based convergence criterion. The described iterative procedure is illustrated in Fig.~\ref{fig:flowchart}. It has been implemented in the open-source FE software GetDP~\cite{dular_1999} combined with additional Python code.
\section{DC-Biased HomHBFEM}\label{sec:method_dc_biased}
\subsection{Naive Extension to DC-Biased Excitation}\label{sec:method_naive_dcbiased_homhbfem}
The first important distinction of the dc-biased case is that we always have to consider both even and odd harmonics. The main difference to the purely harmonic case, however, lies in the way the HBFEM is combined with homogenization.

The naive way of including a dc bias in the analysis is the following. The system of equations remains the same as in Eq. \eqref{eq:homhbfem}, the only difference is that now we also have an equation for $n = 0$ and in this equation, we cannot assemble the $\mathbf{K}_{\nutensor_0 \left(n\wf \right)}$ using the formula given in Eq.~\eqref{eq:tensor_reluctivity} directly because the reluctivity tensor is not defined for $\omega = 0$ as $\delta$ would become infinite. Instead, we use the simple homogenization formula 
\begin{equation}
    \begin{split}
		\underline{\nutensor} &=  \frac{1}{\left( 1 - \gamma \right) \nu_{\mathrm{ins}}^{-1} + \gamma \nu^{-1} }
		\begin{bmatrix}
			1 & 0 & 0 \\ 0 & 1 & 0 \\ 0 & 0 & 0 
		\end{bmatrix} \\  & + \gamma \nu + \left(1 - \gamma\right)\nu_{\mathrm{ins}} \begin{bmatrix}
			0 & 0 & 0 \\ 0 & 0 & 0 \\ 0 & 0 & 1
		\end{bmatrix} \label{eq:tensor_reluctivity_simple}
	\end{split}
\end{equation}
for $n = 0$~\cite{silva_1995}. For $n \neq 0$ we continue to use Eq.~\eqref{eq:tensor_reluctivity} with the modifications for non-negligible insulation thickness and perpendicular magnetic flux. This naive extension of the HomHBFEM to the dc-biased case is straightforward since Eq.~\eqref{eq:tensor_reluctivity_simple} is the limit of  Eq.~\eqref{eq:tensor_reluctivity} combined with Eqs.~\eqref{eq:nu_xy} and \eqref{eq:nu_z} for $\omega \rightarrow 0$. However, as we will demonstrate, it performs poorly in practice. Hence, a more fundamental adaptation of the method is needed. To that end, we start by modifying the homogenization itself.

\subsection{Modified Homogenization Technique}~\label{sec:homogenization_modified}
The derivation of the original homogenization technique is given in~\cite{dular_2003aa}. It is based on the analytical solution of the MQS problem in one lamination, assuming linear material properties and assuming that width and height of the lamination are much larger than its thickness, such that a 1-D analysis suffices. Choosing the coordinate system with its origin in the center of the lamination, the analytical solution for the phasors of the magnetic field strength $\underline{H}$ and the magnetic flux density $\underline{B}$ read
\begin{align}
    \underline{H}(z) &= \underline{H}_{\mathrm{s}} \frac{\cosh\left(k z\right)}{\cosh\left(k \frac{d}{2}\right)} \label{eq:h_cosh},\\
    \underline{B}(z) &= \underline{B}_{\mathrm{s}} \frac{\cosh\left(k z\right)}{\cosh\left(k \frac{d}{2}\right)} \label{eq:b_cosh},
\end{align}
where $\underline{H}_{\mathrm{s}}$ and $\underline{B}_{\mathrm{s}}$ are the magnetic field strength and magnetic flux density at the lamination's surface, i.e., at $z = \pm \frac{d}{2}$~\cite{stoll_1974}. The main idea in~\cite{dular_2003aa} is to consider the laminated yoke or core as a source current region where the source current density is given by the analytical result for the induced eddy current density which is computed from Eq.~\eqref{eq:h_cosh} via Amp\`ere's law. 

To adapt this homogenization better to the nonlinear case, one could try to replace Eqs.~\eqref{eq:h_cosh} and~\eqref{eq:b_cosh} with an analytical solution for the nonlinear MQS problem. Such 1-D nonlinear solutions for the magnetic field in a lamination are available, but only under strong assumptions. Namely, in~\cite{mayergoyz_1998aa} an exact analytical solution is derived under the following assumptions: 

\begin{enumerate}
    \item  The $B$--$H$ curve is given by a power law, i.e., \mbox{$H = \left( \frac{B}{k} \right)^n$}.
    \item The magnetic field strength is circularly polarized, i.e., it features $x$- and $y$-components that are $\SI{90}{\degree}$ phase-shifted with respect to each other.
\end{enumerate}
 The nonlinear analytical solution derived under these assumptions is provided in Appendix~\ref{appendix_a}. 

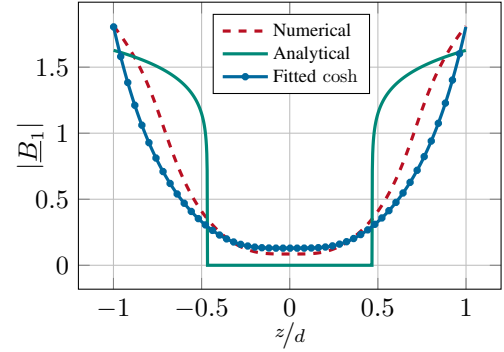
\begin{figure}[t]
\centering
\begin{tikzpicture}
    \begin{axis}[
        width=7.2cm,
        height=5.4cm,
        grid=both,
        xlabel={$\nicefrac{z}{d}$},
        ylabel={$|\underline{B}_1|$},
        xlabel style={yshift=5pt},
        ylabel style={yshift=-10pt},
        legend style={at={(0.69,0.68)}, anchor=south east,  nodes={scale=0.7, transform shape},},
        legend cell align={left},
         x filter/.code={\pgfmathparse{2000*\pgfmathresult}},
    ]

    \addplot[very thick, darkred, dashed] table[
        x index=0,
        y index=1,
        col sep=space,  
        header=false,   
    ] {./figures/results/analytic/b1_mag_num_nl_vs_mayergoyz_vs_fit.txt};
    \addlegendentry{Numerical};

    \addplot[very thick, darkemerald] table[
        x index=0,
        y index=2,
        col sep=space,
        header=false,
    ] {./figures/results/analytic/b1_mag_num_nl_vs_mayergoyz_vs_fit.txt};
    \addlegendentry{Analytical};

    \addplot[very thick, darkblue_two, mark=*,
    mark options={scale=0.4, solid},
    mark repeat={20}] table[
        x index=0,
        y index=3,
        col sep=space,
        header=false,
    ] {./figures/results/analytic/b1_mag_num_nl_vs_mayergoyz_vs_fit.txt};
    \addlegendentry{Fitted $\cosh$};

    \end{axis}
\end{tikzpicture}
\caption{First harmonic of the magnetic flux density in a lamination with $d = \SI{1}{\milli\meter}$ at $f = \SI{1}{\kilo\hertz}$. The numerical simulation uses a Brauer curve for the $B$--$H$ curve, the analytical solution assumes a power law fitted to the Brauer curve.}\label{fig:mayergoyz_solution} 
\end{figure}

Comparing this solution to a nonlinear \mbox{1-D} FE simulation quickly reveals that the analytical solution is not very useful if any of the two above-mentioned assumptions are not met. Fig.~\ref{fig:mayergoyz_solution} shows the magnitude of the first harmonic of the magnetic flux density in a lamination computed with a nonlinear \mbox{1-D} FE simulation using a Brauer model~\cite{brauer_1975} as the $B$-$H$ curve compared to the analytical solution. Additionally, we show the hyperbolic cosine Ansatz in Eq.~\eqref{eq:b_cosh} fitted to the numerical solution. Clearly, the hyperbolic cosine yields a much better approximation than the analytical solution.

Therefore, we stick to the assumption that the magnetic field strength and magnetic flux density can be --- at least approximately --- described by a hyperbolic cosine. However, we adapt this Ansatz by modifying the skin depth $\delta$. We no longer assume $\delta = \sqrt{\frac{2\nu}{\omega\sigma}}$, but use it as a free parameter which we will later determine separately for $\underline{H}$ and $\underline{B}$ in a way that accounts for the ferromagnetic saturation.

Hence, we use as an Ansatz 
\begin{equation}\label{eq:H_nonlin_ana}
    \underline{H}(z) = \underline{H}_{\mathrm{s}} \frac{\cosh\left( \kH z\right)}{\cosh\left(\kH \frac{d}{2}\right)}
\end{equation}
with $\kH = \frac{1 + \jmath}{\deltaH}$ and 
\begin{equation}\label{eq:B_nonlin_ana}
    \underline{B}(z) = \underline{B}_{\mathrm{s}} \frac{\cosh\left( \kB z\right)}{\cosh\left(\kB \frac{d}{2}\right)}
\end{equation}
with $\kB = \frac{1 + \jmath}{\deltaB}$. 

Now the derivation outlined in \cite{dular_2003aa} is repeated, using these modified Ansatz functions. The detailed derivation is provided in Appendix~\ref{appendix_homogenization}. Here, we only give the final result for the homogenized reluctivity which we use for the $x$- and $y$-components of the reluctivity tensor:
\begin{multline}\label{eq:modified_hom}
\nutensor_{xy} =
\frac{\nu \kB^2 d}{8 \sinh^2\!\left(\kB \frac{d}{2}\right)}
\left[
  \frac{\sinh\!\left( \kB d\right)}{\kB} + d
\right] \\
-\jmath \frac{d \nu^2 \kH^4}{8 \sigma \omega \sinh^2\!\left(\kH \frac{d}{2}\right)}
\left[
  \frac{\sinh(\kH d)}{\kH} - d
\right].
\end{multline}
We note that if we set $\kH = \kB = \frac{1 + j}{\delta}$ with $\delta = \sqrt{\frac{2\nu}{\omega \sigma}}$, we recover the original formula given in Eq.~\eqref{eq:tensor_reluctivity}. 
\subsection{Refined Extension to DC-Biased Excitation}\label{sec:method_final}
The refined version of the dc-biased HomHBFEM is constructed as follows. As in the naive approach, in Eq.~\eqref{eq:homhbfem}, we assemble the stiffness matrix  $\mathbf{K}_{\nutensor_0 \left(n\wf \right)}$ for $n=0$ using the reluctivity tensor in Eq.~\eqref{eq:tensor_reluctivity_simple}. However, for $n \neq 0$, the stiffness matrix is no longer constructed with the formula of the original homogenization technique, but instead with the one given by Eq.~\eqref{eq:modified_hom}.

To evaluate $\nu$ and $\kB$ in Eq.~\eqref{eq:modified_hom} during the iterative solution via successive substitution, we again use the dc component of the reluctivity, i.e., we set $\nu = \nu_{0}^{i}$ and $\kB = \frac{1+j}{\deltaB}$ with $\deltaB = \sqrt{\frac{2\nu_{0}^{i}}{\sigma n \wf}}$. 

On the other hand, $\kH = \frac{1 + j}{\deltaH}$ is determined from a precomputed look-up table, which we create from a 1-D FE simulation of a lamination for each frequency of interest. The table contains the value of $\deltaH$ at a given level of ferromagnetic saturation, represented by the temporal maximum of the flux-density averaged across the lamination thickness $\overline{B}_{\mathrm{max}} =\max_{t}\left(\frac{1}{d}\int_{-d/2}^{d/2} B(z,t) \d z \right)$.

The pre-computation of $\deltaH$ is conducted based on the averaged eddy current loss density $\overline{p}_{\mathrm{e}}$ in the lamination. In the linear case, one can swiftly derive~\cite{lammeraner_1966}
\begin{equation}\label{eq:loss_density_ana}
    \overline{p}_{\mathrm{e}} = \frac{1}{2\sigma d}\int_{-\frac{d}{2}}^{\frac{d}{2}} |\underline{J}(z)|^2 \d\,z  = \frac{H_{0}^2}{\sigma d \delta} \frac{\sinh\left(\frac{d}{\delta} \right) - \sin\left(\frac{d}{\delta}\right)}{\cosh\left( \frac{d}{\delta}\right) + \cos\left(\frac{d}{\delta} \right)  },
\end{equation}
where $\underline{J}(z) = -\frac{\d\underline{H}(z) }{\d z}$ is the induced eddy current density.
To determine $\deltaH$, we equate the expression in \eqref{eq:loss_density_ana} with the loss density computed numerically from the nonlinear 1-D FE simulation of the lamination and solve for $\delta$.

In each iteration of the dc-biased HomHBFEM, we then only need to compute the maximum magnetic flux densities in each element and retrieve the corresponding value for $\deltaH$ from the look-up table.

Within the developed framework, different variants are possible. For example, one could also pre-compute look-up tables for $\deltaB$ and $\deltaH$ by fitting the Ansatz functions in Eq.~\eqref{eq:H_nonlin_ana} and Eq.~\eqref{eq:B_nonlin_ana} to the numerical solutions for the first harmonic of $B$ and $H$ or one could determine $\deltaB$ based on the magnetic energy and $\deltaH$ based on the losses. However, the loss-based approach outlined above performed best for our test cases and thus we will focus on this version.
\section{Verification}\label{sec:verification}
\begin{figure}[t]
	\centering
	\begin{minipage}[t]{0.2\textwidth}
		\centering
		\includegraphics[width=0.7\textwidth]{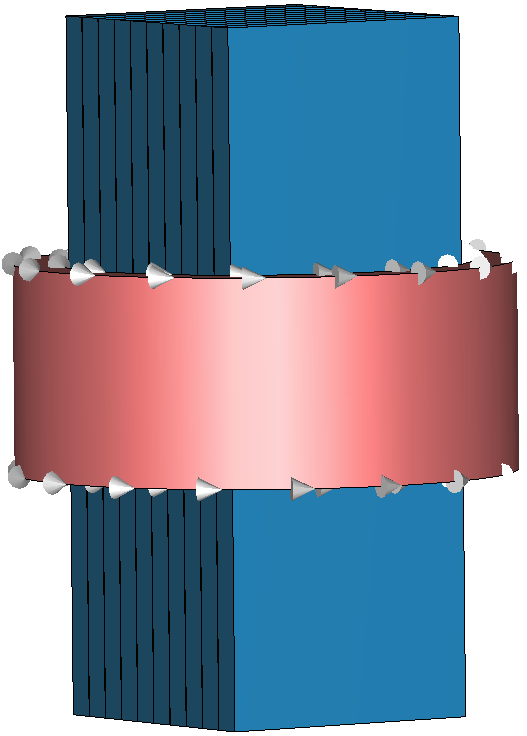}
		\vfill
		\caption{3-D model of the laminated inductor.}
		\label{fig:inductor_model}
	\end{minipage}
	\hspace{0.03\textwidth}
	\begin{minipage}[t]{0.2\textwidth}
		\centering
		\resizebox{1.1\textwidth}{!}{%
				\begin{tikzpicture}[scale = 0.5]
		\begin{axis}[
			xlabel = {$H\,(\si{\kilo\ampere/\meter})$},
			xlabel style={ yshift = 0pt, font=\large},
			ylabel= {$B\,(\si{\tesla})$},
			ylabel style={ yshift = -15pt, font=\large},
			grid = both,
			minor grid style = {gray!15},
			minor grid style = {gray!15},
			legend style={at={(0.05,0.15)},anchor = west,nodes={scale=1.0, transform shape}},
			x filter/.code={\pgfmathparse{0.001*\pgfmathresult}},
			]
			\addplot[line width = 1.5pt, color=darkblue_two] table {./figures/models/BHCurve_Brauer.txt};
		\end{axis}
	\end{tikzpicture}%
		}
		\vfill
		\caption{$B$-$H$ curve based on Brauer model.}
		\label{fig:BH_curve}
	\end{minipage}
\end{figure}

We investigate the 3-D laminated inductor model shown in Fig.~\ref{fig:inductor_model} and we will compare our results with the dc-biased HomHBFEM to a transient reference simulation in CST Studio Suite\textsuperscript{\textregistered}, where the mesh resolves the laminates and the skin depth at the given fundamental frequency.

The model consists of 10 laminations with a thickness of $\SI{0.5}{\milli\meter}$ each. The width of each of the lamination sheets is $\SI{5}{\milli\meter}$ and their height is $\SI{12.5}{\milli\meter}$. For the conductivity of the lamination sheets, we use $\sigma = \SI{10.4} {\mega\siemens\per\meter}$. The stacking factor is $\gamma = \SI{98.5}{\percent}$.

For the magnetization curve, we use a modified version of the Brauer model \cite{brauer_1975}. With $H = | \vec{H} |$ and  $B = | \vec{B} |$, the Brauer curve for the nonlinear material relation is given by
\begin{equation}
	H\left( B \right) = \left( k_1 \mathrm{e}^{k_2 B^2}+ k_3 \right)B, 
\end{equation}
where $k_1,k_2,k_3 \in \mathbb{R}$ are parameters that must be fitted for given measurement data. Given the typical shape of nonlinear $B$-$H$ curves, it is clear that $k_1,k_2 > 0$.

The Brauer model is well-established but it has some shortcomings, one of which is the fact that the resulting reluctivity is unbounded, i.e., 
\begin{equation}
	\lim_{B \to \infty} \nu \left(B \right) = \lim_{B \to \infty}  \frac{H\left( B\right)}{B} = \infty.
\end{equation}
This is physically incorrect because in reality, the reluctivity tends towards the reluctivity of vacuum $\nu_{\mathrm{vac}}$. Furthermore, the fact that the reluctivity tends towards infinity for large $B$ is also numerically problematic, since it can lead to instabilities. Therefore, we modify the Brauer model to
\begin{align}
	\widetilde{H}(B) = 
	\begin{cases}
		( k_1 \mathrm{e}^{k_2 B^2}+ k_3 )B &, \, B \leq B_{\mathrm{s}},  \\
		( k_1 \mathrm{e}^{k_2 B_{\mathrm{s}}^2} + k_3) B_{\mathrm{s}}  + \nu_{\mathrm{vac}}\left(B - B_{\mathrm{s}} \right) &, \, B > B_{\mathrm{s}},\\
	\end{cases} 
\end{align}
where $B_\mathrm{s}$ is the saturation flux density, which we define by $\frac{\d H}{\d B} |_{B = B_{\mathrm{s}}} = \nu_{\mathrm{vac}}$. In this way, we guarantee that the modified $B$-$H$ characteristic is a $C^1$ curve and fulfills 
\begin{equation}
	\lim_{B \to \infty} \nu\left(B\right) = \lim_{B \to \infty}  \frac{\widetilde{H}\left( B\right)}{B} = \nu_{\mathrm{vac}}.
\end{equation}
A similar modification of the Brauer model is proposed in~\cite{hulsmann_2014}. We use ${k_1 = \SI{3.8}{\meter\per\henry}}$, ${k_2 = \SI{2.17}{\tesla^{-2}}}$ and ${k_3 = \SI{396.2}{\meter\per\henry}}$, which are typical values for cold rolled steel, taken from \cite{brauer_1975}. The corresponding $B$-$H$ curve is plotted in Fig.~\ref{fig:BH_curve}.

We will analyze the approximation quality of our method for four different scenarios in terms of the source current. To that end, we will investigate the magnetic energy in the core and the time-averaged eddy current losses. 

\subsection{Scenario A}
In the first investigated scenario, which we refer to as Scenario A, the source current is 
\begin{equation}
	I_{\mathrm{s}}(t) = \SI{562.5}{\ampere} + (\SI{112.5}{\ampere})\cos\left(\wf t\right),
\end{equation}
i.e., the excitation contains a dc bias and a fundamental component whose amplitude corresponds to a fifth of the biasing current. Note that we will refer to the period of the fundamental component as $T_{\mathrm{f}}$ and to the corresponding frequency as $\ff$. We will investigate frequencies between $\ff = \SI{50}{\hertz}$ and $\ff = \SI{10}{\kilo\hertz}$. Figure~\ref{fig:scenario_a_field_profile_50Hz_10kHz} illustrates the magnetic flux density in the laminations for the minimum and maximum considered frequency, giving a sense of the level of ferromagnetic saturation.
\begin{figure}[t]
	\centering
	\resizebox{0.8\linewidth}{!}{%
		\begin{tikzpicture}
\begin{axis}[
    width=10cm,
    height=6cm,
    xlabel={$z$ (\si{\milli\meter})},
    ylabel={$|\vec{B}|$ (\si{\tesla})},
    grid=both,
    grid style={gray!50},
    thick,
    major tick style={semithick},
    tick align=outside,
    ytick = {0,0.25,0.5,0.75,1,1.25,1.5},
    yticklabels = {0,0.25,0.5,0.75,1,1.25,1.5},
]
    \addplot[color = darkred, thick] table [col sep=tab] {figures/results/lam_inductor_scenario_a/b_field_along_z_scenario_a_10kHz_04ms.txt};
     \addplot[color = darkred, thick,dashed] table [col sep=tab] {figures/results/lam_inductor_scenario_a/b_field_along_z_scenario_a_50Hz_80ms.txt};
\end{axis}
\end{tikzpicture}%
	}
	\caption{Absolute value of the magnetic flux density for \mbox{$\ff = \SI{50}{\hertz}$} (dashed) and \mbox{$\ff = \SI{10}{\kilo\hertz}$} (solid) at $t = 4\,T_{\mathrm{f}}$ along a cut through the center of the laminated core in the direction perpendicular to the lamination stack \mbox{(Scenario A)}.}
	\label{fig:scenario_a_field_profile_50Hz_10kHz}
\end{figure}

In this case, we will not only show the results with the refined version of the dc-biased HomHBFEM as explained in Section~\ref{sec:method_final}, but also the results with the naive approach, as explained in Section~\ref{sec:method_naive_dcbiased_homhbfem}. 

Table~\ref{tab:scenarioA_losses} shows the time-averaged eddy current losses for frequencies between $\SI{50}{\hertz}$ and $\SI{10}{\kilo\hertz}$. The refined version of the dc-biased HomHBFEM achieves a good approximation of the losses with a relative error less than $\SI{10}{\percent}$ across all investigated frequency points. We also observe that the refined version of the dc-biased HomHBFEM yields a much better approximation of the reference results than the naive version.
\begin{table}[t]
    \centering
    \caption{Time-averaged eddy current losses in the laminated core (Scenario A)}
    \label{tab:scenarioA_losses}
    \begin{tabular}{l@{\hspace{3pt}}ccc|cc} 
        \toprule
        \toprule
        \multirow{2}{*}{$\ff$\,(\si{\hertz})} 
            & \multicolumn{3}{c}{Power Loss\,(\si{\watt})} 
            & \multicolumn{2}{c}{Rel. Error\,(\si{\percent})} \\ 
        \cmidrule(lr){2-4} \cmidrule(lr){5-6}
         & Reference & Refined & Naive & Refined & Naive \\ 
        \midrule
        50   & $4.79 \cdot 10^{-5}$ & $5.09 \cdot 10^{-5}$ & $5.21 \cdot 10^{-5}$ & $6.3$ & $8.7$ \\ 
        100  & $1.81 \cdot 10^{-4}$ & $1.91 \cdot 10^{-4}$ & $2.02 \cdot 10^{-4}$ & $5.5$ & $11.6$ \\ 
        500  & $3.26 \cdot 10^{-3}$ & $3.29 \cdot 10^{-3}$ & $3.91 \cdot 10^{-3}$ & $0.9$ & $19.9$ \\
        1k & $1.00 \cdot 10^{-2}$ & $1.03 \cdot 10^{-2}$ & $1.22 \cdot 10^{-2}$ & $3.0$ & $22.2$ \\ 
        5k & $1.08 \cdot 10^{-1}$ & $9.98 \cdot 10^{-2}$ & $1.40 \cdot 10^{-1}$ & $7.6$ & $29.6$ \\ 
        10k& $3.01 \cdot 10^{-1}$ & $2.73 \cdot 10^{-1}$ & $3.98 \cdot 10^{-1}$ & $9.3$ & $32.2$ \\
        \bottomrule
        \bottomrule
    \end{tabular}
\end{table}
For all frequencies listed in Table~\ref{tab:scenarioA_losses}, the dc-biased HomHBFEM was applied using the same mesh with $1.1 \cdot 10^5$ DoFs, leading to a simulation time of roughly $90$ minutes per frequency point.

On the other hand, for the transient reference simulation, the mesh must be refined as the frequency is increased to resolve the decreasing skin depth. As a rule of thumb, to determine a fine-enough mesh a-priori, we set the maximum mesh size within the lamination in the transient reference simulation equal to the skin depth $\delta_{\mathrm{in}} = \sqrt{\nicefrac{2\nu_{\mathrm{in}}}{\sigma \wf}}$, where $\nu_{\mathrm{in}}$ is the initial reluctivity of the lamination material, i.e., the reluctivity at zero field. This leads to a transient FE problem with \mbox{$1.9 \cdot 10^{7}$ DoFs} at $\ff = \SI{5}{\kilo\hertz}$ and simulation times of roughly 3 days per period. For the transient reference simulation at $\ff = \SI{10}{\kilo\hertz}$, due to the prohibitive computational effort, we used the same mesh as for $\ff = \SI{5}{\kilo\hertz}$ instead of refining it further. As we can see from Table~\ref{tab:losses_convergence_scenario_a_10kHz_cst}, further refinement is not necessary. In fact, the losses at $\ff = \SI{10}{\kilo\hertz}$ converge with about $5.5 \cdot 10^{6}$ DoFs, corresponding to roughly $\SI{17.5}{\hour}$ of simulation time per period. Since roughly 3 periods need to be simulated for the numerically induced transients to decay, the total time needed for the transient reference simulation at $\ff = \SI{10}{\kilo\hertz}$ is about 2 days. 
\begin{table}[t]
    \centering
    \caption{Convergence of time-averaged eddy current losses in the laminated core at $\ff = \SI{10}{\kilo\hertz}$ (Scenario A)}  \label{tab:losses_convergence_scenario_a_10kHz_cst}  
    \begin{tabular}{cc|cc}
        \toprule
        \toprule
        \multicolumn{2}{c}{Reference} & \multicolumn{2}{c}{HomHBFEM}\\
         \cmidrule(lr){1-2} \cmidrule(lr){3-4}
        \# DoFs & Power Loss (W) &  \# DoFs & Power Loss (W) \\
        \midrule 
         $1.4 \cdot 10^{4}$& $1.14$ & $1.2 \cdot 10^{4}$ &  $0.35$  \\
         $4.8 \cdot 10^{4}$& $0.81$ & $2.5 \cdot 10^{4}$ &  $0.33$ \\
         $1.1 \cdot 10^{5}$ & $0.49$ & $3.3 \cdot 10^{4}$ & $0.30$  \\
         $2.1 \cdot 10^{5}$ & $0.46$ & $4.5 \cdot 10^{4}$ & $0.29$  \\
         $8.6 \cdot 10^{5}$ & $0.34$ & $7.4 \cdot 10^{4}$ & $0.28$  \\
         $2.5 \cdot 10^{6}$ & $0.31$ & $1.1 \cdot 10^{5}$ & $0.27$ \\
          $5.5 \cdot 10^{6}$ & $0.30$ & $2.4 \cdot 10^{5}$ & $0.27$ \\
         $1.9 \cdot 10^{7}$ & $0.30$ & $5.2 \cdot 10^{5}$ & $0.27$   \\
         \bottomrule
         \bottomrule
    \end{tabular}
\end{table}
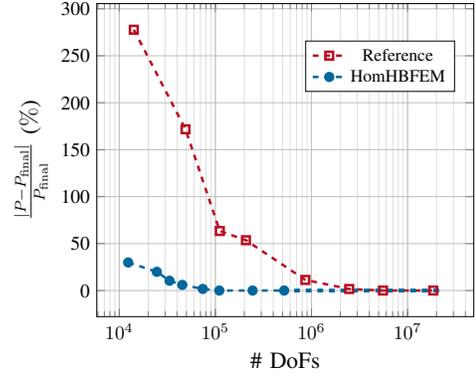
\begin{figure}[t]
	\centering
	\resizebox{0.7\linewidth}{!}{%
		\begin{tikzpicture}
		\begin{semilogxaxis}[
			xlabel = {\# DoFs},
			ylabel ={$\frac{|P - P_\mathrm{final}|}{P_\mathrm{final}}$ (\si{\percent})},
			xlabel style={font=\large},
			ylabel style={font=\large},
			grid = both,
			minor grid style = {gray!25},
			minor grid style = {gray!25},
            ytick={0,50,100,150,200,250,300},
            yticklabels={0,50,100,150,200,250,300},
            legend style={at={(0.95,0.8)},anchor = east,nodes={scale=0.9, transform shape}},
			] 
			\addplot[darkred,dashed, line width=1.3pt, ,mark=square, mark options={solid}] table[x=DoFs, y=Error(percent) ] {./figures/results/lam_inductor_scenario_a/losses_scen_a_10kHz_rel_error_convergence_cst.txt};
            \addlegendentry{Reference};
            \addplot[darkblue_two,dashed, line width=1.3pt, ,mark=*, mark options={solid}] table[x=DoFs, y=Error(percent) ] {./figures/results/lam_inductor_scenario_a/losses_scen_a_10kHz_rel_error_convergence_homhbfem.txt};
            \addlegendentry{HomHBFEM};
            \addplot[darkblue_two,dashed,line width=2pt] coordinates {(5e+5,0) (2.3e+7,0)};
            
		\end{semilogxaxis}
\end{tikzpicture}%
	}
	\caption{Relative deviation from the final result for the time-averaged eddy current losses in the laminated core as a function of the number of DoFs (Scenario A).}
	\label{fig:losses_10kHz_convergence}
\end{figure}

In addition to the convergence of the time-averaged eddy current losses computed with the transient reference simulation, Table~\ref{tab:losses_convergence_scenario_a_10kHz_cst} also shows the convergence of the losses computed with the refined version of the dc-biased HomHBFEM. We observe that in general, the overestimation of the losses due to an insufficiently fine mesh is much more severe with the transient reference simulation than it is with the dc-biased HomHBFEM. Moreover, we can see that with the dc-biased HomHBFEM, the losses converge to their final value with much less DoFs. This is further illustrated in Fig.~\ref{fig:losses_10kHz_convergence}, which shows the relative deviation from the final value for the losses as a function of the number of DoFs for both the transient reference simulation and the dc-biased HomHBFEM. We observe that the number of DoFs required to reach convergence of the losses is reduced by roughly 1.5 orders of magnitude thanks to the dc-biased HomHBFEM.

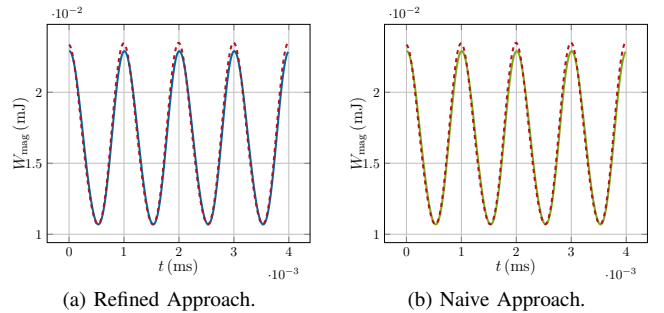
\begin{figure}[b]
	\centering
	\subfloat[Refined Approach.\label{fig:energy_scenario_A_final}]{
		\begin{tikzpicture}[scale=0.51]
			\begin{axis}[
				xlabel = {$t\,(\si{\milli\second})$},
				xlabel style={ yshift = 5pt, font=\large },
				ylabel= {$W_{\mathrm{mag}}\,(\si{\milli\joule})$},
				ylabel style={ yshift = -10pt, font=\large},
				yticklabel style={/pgf/number format/fixed},
				grid = both,
				minor grid style = {gray!15},
				legend style={at={(0.05,0.1)},anchor = west,nodes={scale=0.8, transform shape}},
				y filter/.code={\pgfmathparse{1000*\pgfmathresult}},
				]
				\addplot[line width = 1.5pt, color=darkblue_two]
				table[x index=0, y index=1, header=true]
				{figures/results/lam_inductor_scenario_a/energy_comparison_1000Hz_scenario_a.txt};
				\addplot[line width = 1.5pt, color=darkred, dashed]
				table[x index=0, y index=2, header=true]
				{figures/results/lam_inductor_scenario_a/energy_comparison_1000Hz_scenario_a.txt};
			\end{axis}
		\end{tikzpicture}
	}
	\hspace{0.1cm}
	\subfloat[Naive Approach.\label{fig:energy_scenario_A_naive}]{
		\begin{tikzpicture}[scale=0.51]
			\begin{axis}[
				xlabel = {$t\,(\si{\milli\second})$},
				xlabel style={ yshift = 5pt, font=\large },
				ylabel= {$W_{\mathrm{mag}}\,(\si{\milli\joule})$},
				ylabel style={ yshift = -10pt, font=\large},
				yticklabel style={/pgf/number format/fixed},
				grid = both,
				minor grid style = {gray!15},
				legend style={at={(0.05,0.1)},anchor = west,nodes={scale=0.8, transform shape}},
				y filter/.code={\pgfmathparse{1000*\pgfmathresult}},
				]
				\addplot[line width = 1.5pt, color=darkgreen]
				table[x index=0, y index=1, header=true]
				{figures/results/lam_inductor_scenario_a/energy_comparison_1000Hz_scenario_a.txt};
				\addplot[line width = 1.5pt, color=darkred, dashed]
				table[x index=0, y index=2, header=true]
				{figures/results/lam_inductor_scenario_a/energy_comparison_1000Hz_scenario_a_naive_approach.txt};
			\end{axis}
		\end{tikzpicture}
	}
	\caption{Magnetic energies in the laminated core. DC-biased \mbox{HomHBFEM} (solid) vs. transient reference results (dashed) at $\ff = \SI{1}{\kilo\hertz}$ (Scenario A).}
	\label{fig:energy_scenario_A}
\end{figure}

Next, we will investigate the magnetic energies in the laminated core. Figure~\ref{fig:energy_scenario_A} shows the time signal of the magnetic energy computed with the two versions of the dc-biased HomHBFEM at $\ff=\SI{1}{\kilo\hertz}$ compared to the transient reference simulation. Table~\ref{tab:energy_scenario_A} lists the average and maximum relative errors for the magnetic energies again for both the refined and the naive version of the dc-biased HomHBFEM. As we can see from both Fig.~\ref{fig:energy_scenario_A} and Table~\ref{tab:energy_scenario_A}, the two version of the dc-biased HomHBFEM yield essentially the same results for the magnetic energy, i.e., both yield a good approximation with a maximum relative error less than $\SI{10}{\percent}$ at $\ff = \SI{10}{\kilo\hertz}$. 
Given the much better accuracy of the eddy current losses computed with the refined version of the dc-biased HomHBFEM, we will from now on focus exclusively on this version and will refer to it simply as dc-biased HomHBFEM.

\begin{table}[t]
    \centering
    \caption{Maximum and time-averaged relative error of the magnetic energy in the laminated core (Scenario A)}
    \label{tab:energy_scenario_A}
    \begin{tabular}{lcc c cc} 
        \toprule
        \toprule
        \multirow{2}{*}{$\ff$\,(\si{\hertz})} & \multicolumn{2}{c}{Refined} & & \multicolumn{2}{c}{Naive} \\ 
        \cmidrule{2-3} \cmidrule{5-6}
         & max. error & av. error & & max. error & av. error\\ 
        \midrule
        50 & $\SI{2.6}{\percent}$ & $\SI{1.1}{\percent}$ & & $\SI{2.6}{\percent}$ & $\SI{1.1}{\percent}$ \\
        100 & $\SI{2.1}{\percent}$ & $\SI{1.1}{\percent}$ & & $\SI{2.2}{\percent}$ & $\SI{1.1}{\percent}$ \\
        500 & $\SI{2.4}{\percent}$ & $\SI{1.0}{\percent}$ & & $\SI{2.6}{\percent}$ & $\SI{1.3}{\percent}$ \\
        1k & $\SI{3.6}{\percent}$ & $\SI{2.1}{\percent}$ & & $\SI{3.8}{\percent}$ & $\SI{2.2}{\percent}$ \\
        5k & $\SI{5.5}{\percent}$ & $\SI{2.5}{\percent}$ & & $\SI{6.8}{\percent}$ & $\SI{2.6}{\percent}$ \\
        10k & $\SI{8.3}{\percent}$ & $\SI{4.9}{\percent}$ & & $\SI{8.5}{\percent}$ & $\SI{5.2}{\percent}$ \\
        \bottomrule
        \bottomrule
    \end{tabular}
\end{table}
All in all, we conclude that for this first investigated scenario, the dc-biased HomHBFEM offers a very good tradeoff between accuracy and computational effort, predicting both the eddy current losses and magnetic energies with relative errors below $\SI{10}{\percent}$, reducing the necessary number of DoFs by 1.5 orders of magnitude, and bringing down simulation times at $\ff=\SI{10}{\kilo\hertz}$ from 2 days to just $90$ minutes. Note that the dc-biased HomHBFEM simulations were performed on a computer with $\SI{64}{\giga\byte}$ RAM and an Intel Core i9-13900HX processor. Due to the large memory requirements and long simulations times, the transient reference simulations were  performed on different contemporary servers with either $\SI{512}{\giga\byte}$ or $\SI{256}{\giga\byte}$ RAM and Intel Xeon processors.
\subsection{Scenario A2}
In the following, we will investigate how the method performs for other excitations.
We begin by increasing the amplitude of the oscillation in the excitation current by a factor of 2 while keeping the same dc component as before. Hence, the excitation current is 
\begin{equation}
	I_{\mathrm{s}}(t) = \SI{562.5}{\ampere} + (\SI{225}{\ampere})\cos\left(\wf t\right).
\end{equation}
We refer to this excitation as Scenario A2. Again, to give a sense of the level of ferromagnetic saturation in this scenario, Fig.~\ref{fig:scenario_a2_field_profile_50Hz_10kHz} shows the magnetic flux density on a cut through the center of the laminations in the $z$-direction perpendicular to the laminations.
\begin{figure}[b]
	\centering
	\resizebox{0.7\linewidth}{!}{%
		 \begin{tikzpicture}
    \begin{axis}[
        width=10cm,
        height=6cm,
        xlabel={$z$ (\si{\milli\meter})},
        ylabel={$|\vec{B}|$ (\si{\tesla})},
        grid=both,
        grid style={ gray!50},
        thick,
        major tick style={semithick},
        tick align=outside,
        ytick = {0,0.25,0.5,0.75,1,1.25,1.5,1.75},
        yticklabels = {0,0.25,0.5,0.75,1,1.25,1.5,1.75},
    ]
        \addplot[color = darkred, thick] table [col sep=tab] {./figures/results/lam_inductor_scenario_a2/field_profile_a2_10kHz_04ms.txt};
        \addplot[color = darkred, thick, dashed] table [col sep=tab] {./figures/results/lam_inductor_scenario_a2/field_profile_a2_50Hz_80ms.txt};
    \end{axis}
  \end{tikzpicture}%
	}
	\caption{Absolute value of the magnetic flux density at \mbox{$\ff = \SI{50}{\hertz}$} (dashed) and \mbox{$\ff = \SI{10}{\kilo\hertz}$} (solid) at  $t = 4\,T_{\mathrm{f}}$ along a cut through the center of the laminated core in the direction perpendicular to the lamination stack (Scenario A2).}
	\label{fig:scenario_a2_field_profile_50Hz_10kHz}
\end{figure}

The comparison of the time-averaged eddy current losses computed with the dc-biased HomHBFEM to those computed from the transient reference simulation is shown in Table~\ref{tab:losses_inductor_scenario_a2}. We observe a good approximation of the losses for frequencies up to $\SI{1}{\kilo\hertz}$ with relative errors \mbox{below $\SI{5}{\percent}$.} At $\ff = \SI{5}{\kilo\hertz}$ and $\ff = \SI{10}{\kilo\hertz}$, the relative error rises to $\SI{13}{\percent}$ and $\SI{20}{\percent}$, respectively. 

\begin{figure}[H]
	\centering
	
	\subfloat[$f_{\mathrm{f}} = \SI{50}{\hertz}$\label{fig:tm_energy_A2_50Hz}]{
		\begin{tikzpicture}[scale=0.51]
			\begin{axis}[
				xlabel = {$t\,(\si{\milli\second})$},
				xlabel style={ yshift = 5pt, font=\large},
				ylabel= {$W_{\mathrm{mag}}\,(\si{\milli\joule})$},
				ylabel style={ yshift = -10pt,font=\large},
				yticklabel style={/pgf/number format/fixed},
				grid = both,
				minor grid style = {gray!15},
				legend style={at={(0.05,0.12)},anchor = west,nodes={scale=0.9, transform shape}},
				y filter/.code={\pgfmathparse{1000*\pgfmathresult}},
				x filter/.code={\pgfmathparse{1000*\pgfmathresult}},
				]
				\addplot[line width = 1.5pt, color=darkblue_two]
				table[x index=0, y index=1, header=true]
				{figures/results/lam_inductor_scenario_a2/energy_comparison_50Hz_scenario_a2.txt};
				\addlegendentry{HomHBFEM};
				\addplot[line width = 1.5pt, color=darkred,dashed]
				table[x index=0, y index=2, header=true]
				{figures/results/lam_inductor_scenario_a2/energy_comparison_50Hz_scenario_a2.txt};
				\addlegendentry{Reference};
			\end{axis}
		\end{tikzpicture}
	}
	\hspace{0.1cm}
	\subfloat[$f_{\mathrm{f}} = \SI{100}{\hertz}$\label{fig:tm_energy_A2_100Hz}]{
		\begin{tikzpicture}[scale=0.51]
			\begin{axis}[
				xlabel = {$t\,(\si{\milli\second})$},
				xlabel style={ yshift = 5pt, font=\large },
				ylabel= {$W_{\mathrm{mag}}\,(\si{\milli\joule})$},
				ylabel style={ yshift = -10pt, font=\large},
				yticklabel style={/pgf/number format/fixed},
				grid = both,
				minor grid style = {gray!15},
				legend style={at={(0.05,0.1)},anchor = west,nodes={scale=0.8, transform shape}},
				y filter/.code={\pgfmathparse{1000*\pgfmathresult}},
				]
				\addplot[line width = 1.5pt, color=darkblue_two]
				table[x index=0, y index=1, header=true]
				{figures/results/lam_inductor_scenario_a2/energy_comparison_100Hz_scenario_a2.txt};
				\addplot[line width = 1.5pt, color=darkred,dashed]
				table[x index=0, y index=2, header=true]
				{figures/results/lam_inductor_scenario_a2/energy_comparison_100Hz_scenario_a2.txt};
			\end{axis}
		\end{tikzpicture}
	}
	
	\vspace{0.25cm}
	
	\subfloat[$f_{\mathrm{f}} = \SI{500}{\hertz}$]{
		\begin{tikzpicture}[scale=0.51]
			\begin{axis}[
				xlabel = {$t\,(\si{\milli\second})$},
				xlabel style={ yshift = 5pt, font=\large },
				ylabel= {$W_{\mathrm{mag}}\,(\si{\milli\joule})$},
				ylabel style={ yshift = -10pt, font=\large},
				yticklabel style={/pgf/number format/fixed},
				grid = both,
				minor grid style = {gray!15},
				legend style={at={(0.05,0.1)},anchor = west,nodes={scale=0.8, transform shape}},
				y filter/.code={\pgfmathparse{1000*\pgfmathresult}},
				]
				\addplot[line width = 1.5pt, color=darkblue_two]
				table[x index=0, y index=1, header=true]
				{figures/results/lam_inductor_scenario_a2/energy_comparison_500Hz_scenario_a2.txt};
				\addplot[line width = 1.5pt, color=darkred,dashed]
				table[x index=0, y index=2, header=true]
				{figures/results/lam_inductor_scenario_a2/energy_comparison_500Hz_scenario_a2.txt};
			\end{axis}
		\end{tikzpicture}
	}
	\hspace{0.1cm}
	\subfloat[$f_{\mathrm{f}} = \SI{1}{\kilo\hertz}$\label{fig:energy_A2_1000Hz}]{
		\begin{tikzpicture}[scale=0.51]
			\begin{axis}[
				xlabel = {$t\,(\si{\milli\second})$},
				xlabel style={ yshift = 5pt, font=\large },
				ylabel= {$W_{\mathrm{mag}}\,(\si{\milli\joule})$},
				ylabel style={ yshift = -10pt, font=\large},
				yticklabel style={/pgf/number format/fixed},
				grid = both,
				minor grid style = {gray!15},
				legend style={at={(0.05,0.1)},anchor = west,nodes={scale=0.8, transform shape}},
				y filter/.code={\pgfmathparse{1000*\pgfmathresult}},
				]
				\addplot[line width = 1.5pt, color=darkblue_two]
				table[x index=0, y index=1, header=true]
				{figures/results/lam_inductor_scenario_a2/energy_comparison_1000Hz_scenario_a2.txt};
				\addplot[line width = 1.5pt, color=darkred,dashed]
				table[x index=0, y index=2, header=true]
				{figures/results/lam_inductor_scenario_a2/energy_comparison_1000Hz_scenario_a2.txt};
			\end{axis}
		\end{tikzpicture}
	}
	
	\caption{Magnetic energies in the laminated core. DC-biased \mbox{HomHBFEM} vs. transient reference results (Scenario A2).}
	\label{fig:energy_lam_inductor_scenario_a2}
\end{figure}
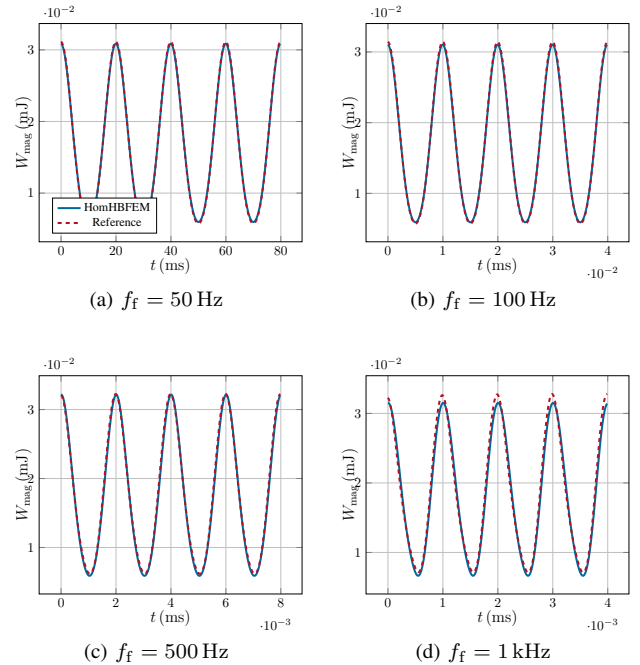

With regard to the magnetic energy in the laminated core, the situation is similar as for the losses. As we can see from Fig.~\ref{fig:energy_lam_inductor_scenario_a2}, the energies at frequencies up to $\SI{1}{\kilo\hertz}$ are well approximated, but at $\ff = \SI{5}{\kilo\hertz}$ and $\SI{10}{\kilo\hertz}$, the error is significantly increased, see Table~\ref{tab:energy_scenario_A2}. 

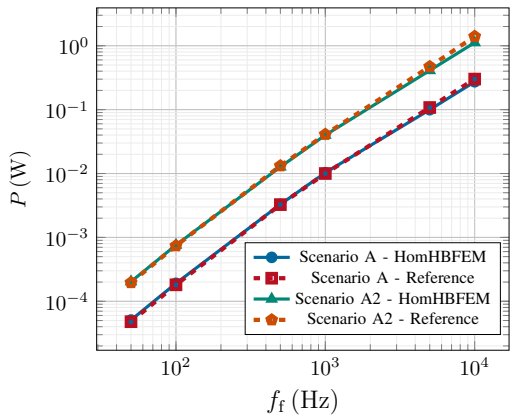
\begin{figure}[H]
    \centering
    \begin{tikzpicture}[scale=0.8]
    		\begin{loglogaxis}[
    			xlabel = {$\ff\,(\si{\hertz})$},
    			ylabel ={$P\,(\si{\watt})$},
    			xlabel style={font=\large},
    			grid = both,
    			minor grid style = {gray!15},
    			minor grid style = {gray!15},
    			legend style={at={(0.98,0.18)},anchor = east,nodes={scale=0.8, transform shape}},
    			]
    			\addplot[darkblue_two,mark=*,line width=1.5pt] table[x=f(Hz), y=HomHBFEM] {figures/results/lam_inductor_scenario_a/losses_scenario_a.txt};
    			\addlegendentry{Scenario A - HomHBFEM};
    			\addplot[darkred,dashed,line width=2pt,mark=square, mark options={solid}] table[x=f(Hz), y=Reference] {figures/results/lam_inductor_scenario_a/losses_scenario_a.txt};
    			\addlegendentry{Scenario A - Reference};
    
                \addplot[darkemerald,mark=triangle*,line width=1.5pt] table[x=f(Hz), y=HomHBFEM] {figures/results/lam_inductor_scenario_a2/losses_scenario_a2.txt};
    			\addlegendentry{Scenario A2 - HomHBFEM};
    			\addplot[darkorange,dashed,line width=2pt,mark=pentagon, mark options={solid}] table[x=f(Hz), y=Reference] {figures/results/lam_inductor_scenario_a2/losses_scenario_a2.txt};
    			\addlegendentry{Scenario A2 - Reference};
    		\end{loglogaxis}
    	\end{tikzpicture}
        \caption{Time-averaged eddy current losses for Scenario A and A2 as a function of the fundamental frequency.}\label{fig:losses_overview_scenario_a_a2}
\end{figure}

All in all, for Scenario A2, the dc-biased HomHBFEM yields a very good approximation of the eddy current losses and the magnetic energies in the laminated core up to $\ff = \SI{1}{\kilo\hertz}$. For the higher frequencies, $\SI{5}{\kilo\hertz}$ and $\SI{10}{\kilo\hertz}$, the approximation quality is significantly decreased but still acceptable, with a maximum relative error of the losses of $\SI{20}{\percent}$ at $\SI{10}{\kilo\hertz}$. To give a quick overview of the results so far, Fig.~\ref{fig:losses_overview_scenario_a_a2} shows the time-averaged eddy current losses in the laminated core for both Scenario A and Scenario A2 computed with the dc-biased HomHBFEM and the transient reference simulation over the full investigated frequency range.
\begin{table}[H]
    \centering
     \caption{Time-averaged eddy current losses in the laminated core (Scenario A2)}\label{tab:losses_inductor_scenario_a2}
    \begin{tabular}{lccc} 
        \toprule
        \toprule
         \multirow{2}{*}{$\ff$\,(\si{\hertz})} & \multicolumn{3}{c}{Power Loss (\si{\watt})}  \\ 
        \cmidrule{2-4} 
        &  HomHBFEM & Reference & Rel. Error\\ 
        \midrule
        50 & $2.03 \cdot 10^{-4}$  & $1.96 \cdot 10^{-4}$ &  $\SI{3.5}{\percent}$  \\ 
        100 & $7.58 \cdot 10^{-4}$  &  $7.39 \cdot 10^{-4}$& $\SI{2.6}{\percent}$\\ 
        500 & $1.27 \cdot 10^{-2}$  & $1.32 \cdot 10^{-2}$ & $\SI{3.8}{\percent}$  \\  
        1k & $4.0 \cdot 10^{-2}$ &  $4.10 \cdot 10^{-2}$ & $\SI{2.4}{\percent}$\\
        5k & $4.10 \cdot 10^{-1}$ &  $4.71 \cdot 10^{-1}$ & $\SI{13.0}{\percent}$  \\
        10k & $1.12$ &  $1.40$  & $\SI{20.0}{\percent}$ \\
        \bottomrule
        \bottomrule
    \end{tabular}
\end{table}

\begin{table}[H]
    \centering
    \caption{Maximum and time-averaged relative error of the magnetic energy in the laminated core (Scenario A2)}
    \label{tab:energy_scenario_A2}
    \begin{tabular}{lcc} 
        \toprule
        \toprule
        $\ff$\,(\si{\hertz}) & max. error & av. error \\ 
        \midrule
        50 & $\SI{3.9}{\percent}$ & $\SI{1.6}{\percent}$ \\
        100 & $\SI{3.2}{\percent}$ & $\SI{1.5}{\percent}$  \\
        500 & $\SI{6.2}{\percent}$ & $\SI{2.8}{\percent}$ \\
        1k & $\SI{8.0}{\percent}$ & $\SI{4.8}{\percent}$\\
        5k & $\SI{13.1}{\percent}$ & $\SI{6.3}{\percent}$ \\
        10k & $\SI{29.2}{\percent}$ & $\SI{14.5}{\percent}$\\
        \bottomrule
        \bottomrule
    \end{tabular}
\end{table}

\subsection{Scenarios B and B2}
We have examined the performance of the method for two more excitations which we refer to as Scenario B and B2. For Scenario B, the source current is
\begin{equation}
	I_{\mathrm{s}}(t) = \SI{1125}{\ampere} + (\SI{112.5}{\ampere})\cos\left(\wf t\right)
\end{equation}
and for Scenario B2 it is 
\begin{equation}
	I_{\mathrm{s}}(t) = \SI{1125}{\ampere} + (\SI{225}{\ampere})\cos\left(\wf t\right),
\end{equation}
i.e., in both scenarios the dc component of the current is twice as high as in Scenarios A and A2 and the fundamental component is the same as in Scenario A and A2, respectively.
Fig.~\ref{fig:scenario_b_b2_field_profile_50Hz} shows the absolute value of the magnetic flux density for both scenarios at $\ff = \SI{50}{\hertz}$ on the same cut as in Figs.~\ref{fig:scenario_a_field_profile_50Hz_10kHz} and~\ref{fig:scenario_a2_field_profile_50Hz_10kHz}. We observe that the outer laminations are now almost fully saturated, such that the increase in the amplitude of  the source current's fundamental component barely changes the magnetic flux density anymore.

The findings for these two scenarios are qualitatively similar as for Scenario A2, i.e., the dc-biased HomHBFEM yields a good approximation of losses and magnetic energies up to $\ff = \SI{1}{\kilo\hertz}$, but for $\ff = \SI{5}{\kilo\hertz}$ and $\ff = \SI{10}{\kilo\hertz}$, the errors are significantly increased. However, we observe that the stronger ferromagnetic saturation of the laminates in Scenarios B and B2 results in a much larger increase of the relative errors at higher frequencies than what we found for Scenario A2. 

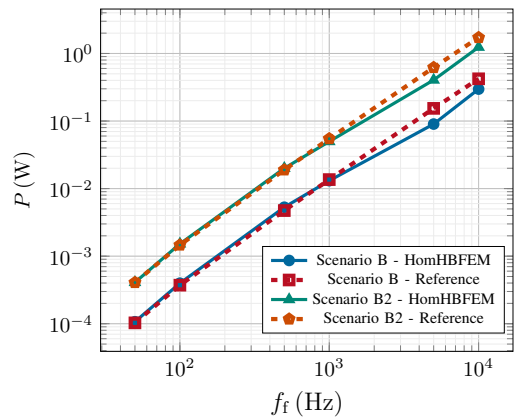
\begin{figure}[t]
\centering
\begin{tikzpicture}[scale=0.8]
        \centering
		\begin{loglogaxis}[
			xlabel = {$\ff\,(\si{\hertz})$},
			ylabel ={$P\,(\si{\watt})$},
			xlabel style={font=\large},
			grid = both,
			minor grid style = {gray!15},
			minor grid style = {gray!15},
			legend style={at={(0.98,0.18)},anchor = east,nodes={scale=0.75, transform shape}},
			]
            \addplot[darkblue_two,mark=*,line width=1.5pt] table[x=f(Hz), y=HomHBFEM] {figures/results/lam_inductor_scenario_b/losses_scenario_b.txt};
			\addlegendentry{Scenario B - HomHBFEM};
			\addplot[darkred,dashed,line width=2pt,mark=square, mark options={solid}] table[x=f(Hz), y=Reference] {figures/results/lam_inductor_scenario_b/losses_scenario_b.txt};
			\addlegendentry{Scenario B - Reference};
             \addplot[darkemerald,mark=triangle*,line width=1.5pt] table[x=f(Hz), y=HomHBFEM] {figures/results/lam_inductor_scenario_b2/losses_scenario_b2.txt};
			\addlegendentry{Scenario B2 - HomHBFEM};
			\addplot[darkorange,dashed,line width=2pt,mark=pentagon, mark options={solid}] table[x=f(Hz), y=Reference] {figures/results/lam_inductor_scenario_b2/losses_scenario_b2.txt};
			\addlegendentry{Scenario B2 - Reference};
		\end{loglogaxis}
	\end{tikzpicture}
    \caption{Time-averaged eddy current losses for Scenario B and B2 as a function of the fundamental frequency.}\label{fig:losses_overview_b_b2}
\end{figure}

The detailed numerical results, i.e., the tables with the time-averaged eddy current losses and the relative errors of the magnetic energies are provided in Appendix~\ref{appendix_b}. For brevity, we show here only Fig.~\ref{fig:losses_overview_b_b2}, which summarized the approximation of the time-averaged eddy current losses by the dc-biased HomHBFEM for both Scenario B and B2. Clearly, we see that the dc-biased HomHBFEM underestimates the losses at $\ff = \SI{5}{\kilo\hertz}$ and $\ff = \SI{10}{\kilo\hertz}$ significantly. 

All in all, for Scenarios B and B2, which represent challenging test cases with a highly saturated laminated core, the dc-biased HomHBFEM only yields acceptable results up to roughly $\ff = \SI{1}{\kilo\hertz}$. At $\ff = \SI{5}{\kilo\hertz}$ and $\ff = \SI{10}{\kilo\hertz}$, the errors in the magnetic energy and the time-averaged eddy current losses become too large. 
\begin{figure}[t]
	\centering
	\resizebox{0.8\linewidth}{!}{%
		     \begin{tikzpicture}
        \begin{axis}[
            width=10cm,
            height=6cm,
            xlabel={$z$ (\si{\milli\meter})},
            ylabel={$|\vec{B}|$ (\si{\tesla})},
            grid=both,
            grid style={ gray!50},
            thick,
            major tick style={semithick},
            tick align=outside,
            ytick = {0,0.25,0.5,0.75,1,1.25,1.5,1.75},
            yticklabels = {0,0.25,0.5,0.75,1,1.25,1.5,1.75},
        ]
         \addplot[color = darkred, thick] table [col sep=tab] {figures/results/lam_inductor_scenario_b2/field_profile_b2_50Hz_80ms.txt};
          \addplot[color = darkred, thick,dashed] table [col sep=tab] {figures/results/lam_inductor_scenario_b/field_profile_b_50Hz_80ms.txt};
        \end{axis}
      \end{tikzpicture}%
	}
	\caption{Absolute value of the magnetic flux density for $\ff = \SI{50}{\hertz}$ at  $t = 4\,T_{\mathrm{f}}$ along a cut through the center of the laminated core in the direction perpendicular to the lamination stack (Scenarios B and B2). }
	\label{fig:scenario_b_b2_field_profile_50Hz}
\end{figure}

\section{Conclusion}\label{sec:conclusion}
The HomHBFEM combines the HBFEM with a frequency-domain-based homogenization technique to facilitate efficient nonlinear eddy current simulations at elevated frequencies with relatively coarse FE meshes. This paper has developed an extension of the HomHBFEM to include excitation currents with a dc bias. To that end, the original homogenization technique was adapted for better suiting the nonlinear case. The resulting formula for the homogenized reluctivity is evaluated with the help of a look-up table which is computed from a 1-D FE simulation of a lamination and contains the average magnetic flux density in the lamination and the corresponding skin depth.

To assess the performance of the proposed method, we used a 3-D model of a laminated inductor and compared the results of the dc-biased HomHBFEM to fine-mesh transient reference simulations. We investigated a broad frequency range between $\SI{50}{\hertz}$ and $\SI{10}{\kilo\hertz}$ and different levels of magnetic saturation, represented by four different scenarios. In \mbox{Scenario A}, representing low to medium levels of saturation, the dc-biased HomHBFEM yields a good approximation of the time-averaged eddy current losses and the magnetic energies with relative errors below $\SI{10}{\percent}$. At the same time, the method reduces the required number of DoFs at $\ff = \SI{10}{\kilo\hertz}$ by 1.5 orders of magnitude, thus reducing simulation times from 2 days on a contemporary server to $90$ minutes on a standard workstation. 

As the level of saturation was increased through Scenarios A2, B, and B2, the approximation quality at frequencies up to $\SI{1}{\kilo\hertz}$ remained good, but gradually deteriorated at the higher frequencies, $\SI{5}{\kilo\hertz}$ and $\SI{10}{\kilo\hertz}$. Hence, to achieve a good performance for higher levels of saturation at frequencies above $\SI{1}{\kilo\hertz}$, further improvements are necessary.

All in all, the proposed method constitutes a promising approach for dc-biased nonlinear eddy current simulations at elevated frequencies. Since the application we are working on, the FC magnets for PETRA IV at DESY, does not feature high levels of saturation, the presented approach is practical. Nonetheless, future work should focus on improving the performance at high saturation for excitation frequencies in the kilohertz range to broaden the range of applicability.


%

\appendices
\section{Analytical Solution to the Nonlinear MQS Problem in a Lamination}\label{appendix_a}
The analytical solution to the nonlinear MQS problem in a lamination, as derived in~\cite{mayergoyz_1998aa}, yields for the magnitude of the magnetic flux density
\begin{equation}\label{eq:appendix_b}
    |\underline{B}(z)| =
    \begin{cases}
    \mu_{\mathrm{s}} H_{\mathrm{s}} \left(1 - \frac{\nicefrac{d}{2} + z}{z_0} \right)^{\frac{2}{n-1}} , &  -\frac{d}{2} \leq z < -\frac{d}{2} + z_0, \\[6pt]
   0, & -\frac{d}{2} + z_0 \leq z \leq \frac{d}{2} - z_0, \\[6pt]
     \mu_{\mathrm{s}} H_{\mathrm{s}} \left(1 - \frac{\nicefrac{d}{2}-z}{z_0} \right)^{\frac{2}{n-1}}, &  \frac{d}{2} - z_0 < z \leq \frac{d}{2}.
    \end{cases}
\end{equation}
Herein, $\mu_{\mathrm{s}}$ and $H_{\mathrm{s}}$ are the permeability and the magnitude of the magnetic field strength at the lamination's surface, $n > 1$ is the exponent in the power law 
\begin{equation}\label{eq:appendix_power_law}
    H = \left( \frac{B}{k} \right)^n,
\end{equation}
which is assumed as the $B$--$H$ curve in the derivation, and 
\begin{equation}
    z_0 = \frac{\left[ 2n (n + 1) (3n + 1)^2 \right]^{\frac{1}{4}}}{(n-1) \sqrt{\omega \sigma \mu_{\mathrm{s}} }}
\end{equation}
is the total penetration depth of the magnetic field. Note that besides the assumption of the power law in Eq.~\eqref{eq:appendix_power_law}, the second assumption in the derivation of this solution is that the magnetic field strength is circularly polarized, i.e., it features $x$- and $y$-components that are $\SI{90}{\degree}$ phase-shifted with respect to each other. This assumption results in a magnetic field strength whose magnitude only depends on  the spatial coordinate $z$ and not on $t$. Remarkably, this leads to a solution without higher order harmonics, i.e., Eq.~\eqref{eq:appendix_b} describes the first and only harmonic of the magnetic flux density.
\section{Derivation of the Modified Homogenization}\label{appendix_homogenization}
The derivation of the modified homogenization in Section~\ref{sec:homogenization_modified} follows the derivation of the original homogenization in~\cite{dular_2003aa} closely. Nonetheless, we will give a detailed account of the derivation in the following.

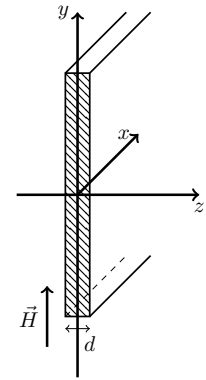
\begin{figure}[b]
	\centering
	\resizebox{0.3\linewidth}{!}{%
			\begin{tikzpicture}
		\draw[->,very thick] (-1,0) -- (2,0) node [below] {$z$};
		\draw[->,very thick] (0,-3) -- (0,3) node [left] {$y$};
		\draw[->,very thick](0,0)--(1 ,1)node [left] {$x$};
		\draw[thick,pattern=north west lines, pattern color=black] (-0.2,-2) rectangle (0.2,2);
		\draw[<->] (-0.2,-2.2) -- (0.2,-2.2)node [below] {$d$};
		\draw[thick] (-0.2,2) -- ++(1,1);
		\draw[thick] (0.2,2) -- ++(1,1);
		\draw[thick] (0.2,-2) -- ++(1,1);
		\draw[dashed] (-0.2,-2) -- ++(1,1);
		\draw[->,very thick] (-0.5,-2.5)--(-0.5,-1.5) node [midway,left] {$\vec{H}$};
	\end{tikzpicture}%
	}
	\caption{Single lamination with coordinate system.}
	\label{fig:appendix_hom_lamination}
\end{figure}

The derivation of the original homogenization is based on the analytical solution of the linear MQS problem in a lamination. We choose the coordinate system as indicated in Fig.~\ref{fig:appendix_hom_lamination}. For simplicity we assume that the magnetic field strength only has a component in $y$-direction, 
but note that the treatment for a magnetic field strength in $x$-direction is completely analogous. Under these assumptions, the magnetic field strength is $\underline{\vec{H}} = \underline{H}(z) \vec{e}_y$ with 
\begin{equation}\label{eq:app_hom_cosh_1}
    \underline{H}(z) = \underline{H}_{\mathrm{s}} \frac{\cosh\left(k z\right)}{\cosh\left(k \frac{d}{2}\right)},
\end{equation}
where $\underline{H}_{\mathrm{s}}$ is the magnetic field strength at the lamination's surface, i.e., at $z = \pm \frac{d}{2}$ and $k = \frac{1+\jmath}{\delta}$. Herein, $\delta$ is the skin depth defined via $\delta = \sqrt{\frac{2\nu}{\sigma\omega}}$.

For the derivation of the modified homogenization, we use
\begin{equation}\label{eq:app_hom_cosh_2}
    \underline{H}(z) = \underline{H}_{\mathrm{s}} \frac{\cosh\left( \kH z\right)}{\cosh\left(\kH \frac{d}{2}\right)},
\end{equation}
where $\kH = \frac{1+\jmath}{\deltaH}$, but $\deltaH$ is now a free parameter, which will later be determined adequately from a nonlinear 1-D simulation of the lamination. 
Via Amp\`eres law, we obtain for the induced eddy current density $\vec{\underline{J}} = \underline{J}(z) \vec{e}_x$ with 
\begin{equation}\label{eq:app_hom_j_1}
    \underline{J}(z) = -\frac{\d \underline{H}(z) }{\d z} =- \kH \underline{H}_{\mathrm{s}} \frac{\sinh\left(\kH z \right)}{\cosh\left(\kH \frac{d}{2} \right)}. 
\end{equation}
We want to rewrite Eq.~\eqref{eq:app_hom_j_1} in terms of the averaged magnetic field flux density $\underline{B}_{\mathrm{a}}$ in the lamination. To that end, we first compute the average magnetic field strength
\begin{align}
    \underline{H}_{\mathrm{a}} &= \frac{1}{d}\int_{-\frac{d}{2}}^{\frac{d}{2}} \underline{H}_{\mathrm{s}} \frac{\cosh\left(\kH z\right)}{\cosh\left(\kH \frac{d}{2}\right)} \d z\\
    &= \frac{2 \underline{H}_{\mathrm{s}} \sinh\left(\kH \frac{d}{2} \right)}{\kH d \cosh\left(\kH \frac{d}{2} \right)}\label{eq:hav_hs}.
\end{align}
From Eq.~\eqref{eq:hav_hs}, we obtain 
\begin{equation}\label{eq:hs_hav}
    \underline{H}_{\mathrm{s}} = \underline{H}_{\mathrm{a}} \frac{\kH d}{2}  \frac{\cosh\left( \kH \frac{d}{2}\right)}{\sinh\left( \kH \frac{d}{2}\right)}
\end{equation}
and inserting Eq.~\eqref{eq:hs_hav} into Eq.~\eqref{eq:app_hom_j_1} together with $H_{\mathrm{a}} = \nu B_{\mathrm{a}}$ yields
\begin{equation}\label{eq:j_z_bav}
    \underline{J}(z) = -\nu \underline{B}_{\mathrm{a}} \frac{\kH^2 d}{2} \frac{\sinh\left( \kH z\right)}{\sinh\left( \kH \frac{d}{2}\right)}.
\end{equation}
Analogously to Eq.~\eqref{eq:app_hom_cosh_2}, the magnetic flux density in the lamination is expressed as
\begin{equation}
    B(z) = \underline{B}_{\mathrm{s}} \frac{\cosh\left(\kB z \right)}{\cosh\left(\kB \frac{d}{2}\right)}
\end{equation}
with $\kB = \frac{1+\jmath}{\deltaB}$ and $\deltaB$ being a free parameter that will later be determined from the nonlinear 1-D simulation. Using the relation between $\underline{B}_{\mathrm{s}}$ and $\underline{B}_{\mathrm{a}}$, which is analogous to Eq.~\eqref{eq:hs_hav}, we obtain
\begin{equation}\label{eq:b_z_bav}
    \underline{B}(z) = \underline{B}_{\mathrm{a}} \frac{\kB d}{2} \frac{\cosh\left(\kB z\right)}{\sinh\left(\kB \frac{d}{2} \right)}.
\end{equation}
The expressions in Eqs.~\eqref{eq:j_z_bav} and~\eqref{eq:b_z_bav} are now introduced into the following weak $\vec{A}^{*}$-formulation in frequency domain. 

Let $\Omega$ be the entire computational domain and $H_{\mathrm{D}}\left(\text{curl};\Omega \right)$ the Sobolev space consisting of all square-integrable vector fields $\underline{\vec{v}} : \Omega \to \mathbb{C}^{3}$ whose (weak) curl is square-integrable and whose tangential components vanish on the Dirichlet part of the boundary. Then, the weak formulation reads: \mbox{Determine ${\AvecComplex \in H_{\mathrm{D}}\left(\text{curl};\Omega \right)}$ such that}
\begin{multline}\label{eq:WF1}
	\int_{\Omega} \left( \nu \nabla \times \AvecComplex\right) \cdot \left(\nabla \times \AvecComplex' \right)\,\d V + \jmath \omega \int_{\Omega_{\mathrm{c}} } \sigma \AvecComplex \cdot \AvecComplex'\,\d V = \\ \int_{\Omega_{\mathrm{s}}} \JsvecComplex \cdot \AvecComplex ' \,\d V \, \, \, \, \, \forall \AvecComplex' \in  H_{\mathrm{D}}\left(\text{curl}; \Omega \right),
\end{multline}
where $\OmegaC \coloneqq \{\vec{r} \in \Omega : \sigma\left(\vec{r}\right) \neq 0 \}$ denotes the eddy current conducting parts of $\Omega$, and $\OmegaS \coloneqq \{\vec{r} \in \Omega : \underline{\vec{J}}_{\mathrm{s}}\left(\vec{r}\right) \neq 0 \}$ the parts with source conductors.

The key idea is to no longer consider the lamination stack as a subset of $\Omega_{\mathrm{c}}$ but as a subset of $\Omega_{\mathrm{s}}$, with the source current prescribed by the analytical result for the induced eddy current density in Eq.~\eqref{eq:j_z_bav}. 

To that end, we rewrite Eq.~\eqref{eq:WF1} as
\begin{multline}\label{eq:WFSplit}
	\int_{\Omega\setminus\Omegals} \left( \nu \nabla \times \underline{\A}\right) \cdot \left(\nabla \times \underline{\A}' \right)\,\d V \\ + \underbrace{\int_{\Omegals} \left( \nu \nabla \times \underline{\A} \right) \cdot \left(\nabla \times \underline{\A}' \right)\,\d V}_{\circled{2}} + j \omega \int_{\OmegaC} \sigma \underline{\A} \cdot  \underline{\A}'\,\d V \\= \int_{\OmegaS\setminus\Omegals} \underline{\vec{J}}_{\mathrm{s}} \cdot \underline{\A}' \d V + \underbrace{\int_{\Omegals} \underline{\vec{J}}_{\mathrm{s}} \cdot \underline{\A}'\,\d V}_{\circled{1}}  \, \, \, \forall \underline{\A}' \in  H_{D}\left(\text{curl}; \Omega \right),
\end{multline}
where $\Omegals$ denotes the lamination stack. 

Let us first focus on term $\circled{1}$. From Faraday's law it follows together with the suitable constitutive equation that inside the lamination stack we have
\begin{equation}\label{eq:test_func_term1}
    \vec{\underline{A}} = -\frac{1}{\jmath\omega\sigma}\vec{\underline{J}} = -\frac{1}{\jmath\omega\sigma} \underline{J}(z)\vec{e}_x 
\end{equation}
with $\underline{J}(z)$ as in Eq.~\eqref{eq:j_z_bav}. Following~\cite{dular_2003aa}, the expression \eqref{eq:test_func_term1} is chosen as a test function. Inserting Eq.~\eqref{eq:test_func_term1} for the test function and Eq.~\eqref{eq:j_z_bav} for the source current density in term $\circled{1}$ yields
\begin{align}\label{Eq:TestFunc}
	\int_{\Omegals} \underline{\vec{J}}_{\mathrm{s}} \cdot \underline{\A}'\,\d V &= -\frac{1}{j\omega \sigma} \int_{\Omegals}   \underline{J}\left(z\right) \underline{J}'\left(z\right)\,\d V \\&= -j \omega \int_{\Omegals} \underline{F}_{J}\left(z \right) \underline{B}_{\mathrm{a}} \underline{B}_{\mathrm{a}}'\,\d V,
\end{align}
with 
\begin{equation}
    \underline{F}_{J}(z) = -\frac{\nu^2 \kH^4 d^2}{4 \omega^2 \sigma} \frac{\sinh^2\left(\kH z\right)}{\sinh^2\left(\kH \frac{d}{2}\right)} 
\end{equation}
To transform the actual lamination stack $\Omegals$ into the homogenized lamination stack $\tilde{\Omega}_{\mathrm{ls}}$, we average $\underline{F}_{J}(z)$ over the thickness of one lamination, i.e., we compute
\begin{align}\label{eq:FJ_Average}
	\underline{F}_{J}^{\mathrm{a}} &= \frac{1}{d} \int_{-d/2}^{d/2}	\underline{F}_{J}\left(z\right)  \d z \\ &= \frac{-d \nu^2 \kH^4}{8 \sigma \omega^2 \sinh^2\left(\kH\frac{d}{2}\right)} \left[\frac{\sinh\left(\kH d \right)}{\kH} - d \right]
\end{align} 
Next, we deal with term $\circled{2}$. Again, we follow~\cite{dular_2003aa} and write
\begin{multline}
	\int_{\Omegals} \left( \nu \nabla \times \underline{\A} \right) \cdot \left( \nabla \times \underline{\A}' \right) \d V \\ = \int_{\Omegals}  \nu \underline{B}\left(z\right) \underline{B}'\left(z\right) \d V  = \int_{\Omegals} \underline{F}_{B} \left( z \right)\underline{B}_{\mathrm{a}} \underline{B}_{\mathrm{a}}' \d V.
\end{multline}
Using Eq.~\eqref{eq:b_z_bav} for $\underline{B}(z)$ leads to 
\begin{equation}\label{eq:Fb_z}
    \underline{F}_{B}(z) = \nu \frac{\kB^2 d^2}{4} \frac{\cosh^2\left( \kB z\right)}{\sinh^2\left(\kB \frac{d}{2}\right)}
\end{equation}
and averaging the expression in Eq.~\eqref{eq:Fb_z} over the lamination thickness yields
\begin{align}
    \underline{F}_{B}^{\mathrm{a}} &= \frac{1}{d} \int_{-d/2}^{d/2}	\underline{F}_{B}\left(z\right)  \d z \\ &= \frac{\nu \kB^2 d}{8 \sinh^2\left(\kB \frac{d}{2}\right)} \left[ \frac{\sinh\left( \kB d\right)}{\kB}  +d\right].
\end{align}
Now we can replace the terms $\circled{1}$ and $\circled{2}$ in Eq.~\eqref{eq:WFSplit} with their homogenized counterparts and combine them into one term for the homogenized lamination stack $\tilde{\Omega}_{\mathrm{ls}}$, i.e., we have
\begin{align}
    \int_{\Omegals} \left( \nu \nabla \times \underline{\A} \right) &\cdot \left(\nabla \times \underline{\A}' \right)\,\d V - \int_{\Omegals} \underline{\vec{J}}_{\mathrm{s}} \cdot \underline{\A}'\,\d V \\ &=
    \int_{\tilde{\Omega}_{\mathrm{ls}}} \left(\underline{F}_{B}^{\mathrm{a}} + \jmath \omega\underline{F}_{J}^{\mathrm{a}} \right) \underline{B}_{\mathrm{a}} \underline{B}_{\mathrm{a}}' \dV.
\end{align}
Since the derivation for a magnetic field strength in $x$-direction is completely analogous, the final result for the $x$- and $y$-components of the reluctivity tensor is thus
\begin{multline}
    \nutensor_{xy} = \underline{F}_{B}^{\mathrm{a}} + \jmath \omega\underline{F}_{J}^{\mathrm{a}} = \\ \frac{\nu \kB^2 d}{8 \sinh^2\left(\kB \frac{d}{2}\right)} \left[ \frac{\sinh\left( \kB d\right)}{\kB}  +d\right] \\-\jmath \frac{d \nu^2 \kH^4}{8 \sigma \omega \sinh^2\left(\kH\frac{d}{2}\right)} \left[\frac{\sinh\left(\kH d \right)}{\kH} - d \right].
\end{multline}
If we set $\kH = \kB = \frac{1 + \jmath}{\delta}$ with $\delta = \sqrt{\frac{2\nu}{\omega \sigma}}$, we recover the original formula presented in~\cite{dular_2003aa}.
\section{Detailed Results for Scenarios B and B2}\label{appendix_b}
\subsection{Scenario B}
\begin{table}[H]
    \centering
     \caption{Time-averaged eddy current losses in the laminated core (Scenario B)}\label{tab:losses_b}
    \begin{tabular}{lccc} 
        \toprule
        \toprule
         \multirow{2}{*}{$\ff$\,(\si{\hertz})} & \multicolumn{3}{c}{Power Loss (\si{\watt})}  \\ 
        \cmidrule{2-4} 
        &  HomHBFEM & Reference & Rel. Error\\ 
        \midrule
        50 & $1.07 \cdot 10^{-4} $  & $1.03 \cdot 10^{-4}$ &  $\SI{3.8}{\percent}$  \\
        100 & $3.98 \cdot 10^{-4}$ &  $3.72 \cdot 10^{-4} $ & $\SI{7.0}{\percent}$\\ 
        500 & $5.30 \cdot 10^{-3}$  & $4.73 \cdot 10^{-4}$ & $\SI{12.1}{\percent}$  \\
        1k & $1.31\cdot 10^{-2}$ &  $1.36 \cdot 10^{-2}$ & $\SI{3.7}{\percent}$ \\ 
        5k & $9.02 \cdot 10^{-2}$ &  $1.54 \cdot 10^{-1}$ & $\SI{41.4}{\percent}$ \\
        10k & $2.97 \cdot 10^{-2}$ &  $4.22 \cdot 10^{-1} $ & $\SI{29.6}{\percent}$ \\
        \bottomrule
        \bottomrule
    \end{tabular}
\end{table}

\begin{table}[H]
    \centering
    \caption{Maximum and time-averaged relative error of the magnetic energy in the laminated core (Scenario B)}
    \label{tab:energy_scenario_B}
    \begin{tabular}{lcc} 
        \toprule
        \toprule
        $\ff$\,(\si{\hertz}) & max. error & av. error \\ 
        \midrule
        50 & $\SI{1.6}{\percent}$ & $\SI{0.9}{\percent}$ \\
        100 & $\SI{2.1}{\percent}$ & $\SI{1.0}{\percent}$  \\
        500 & $\SI{2.2}{\percent}$ & $\SI{1.3}{\percent}$ \\
        1k & $\SI{3.1}{\percent}$ & $\SI{1.4}{\percent}$\\
        5k & $\SI{8.7}{\percent}$ & $\SI{3.9}{\percent}$ \\
        10k & $\SI{7.6}{\percent}$ & $\SI{3.0}{\percent}$\\
        \bottomrule
        \bottomrule
    \end{tabular}
\end{table}

\subsection{Scenario B2}
\begin{table}[H]
    \centering
     \caption{Time-averaged eddy current losses in the laminated core (Scenario B2)}\label{tab:losses_inductor_scenario_b2}
    \begin{tabular}{lccc} 
        \toprule
        \toprule
         \multirow{2}{*}{$\ff$ (\si{\hertz})} & \multicolumn{3}{c}{Power Loss (\si{\watt})}  \\ 
        \cmidrule{2-4} 
        &  HomHBFEM & Reference & Rel. Error\\ 
        \midrule
        50 & $4.09 \cdot 10^{-4}$  & $4.07 \cdot 10^{-4}$ &  $\SI{0.5}{\percent}$  \\
        100 & $1.52\cdot 10^{-3}$ &  $1.46 \cdot 10^{-3}$ & $\SI{4.1}{\percent}$\\ 
        500 & $2.00 \cdot 10^{-2}$  & $1.89 \cdot 10^{-2}$ & $\SI{5.8}{\percent}$  \\
        1k & $4.98 \cdot 10^{-2}$ &  $5.48 \cdot 10^{-2}$ & $\SI{9.1}{\percent}$\\
        5k & $4.05\cdot 10^{-1}$ &  $6.20 \cdot 10^{-1}$ & $\SI{34.7}{\percent}$ \\
        10k & $1.24$ &  $1.72$ & $\SI{27.9}{\percent}$ \\
        \bottomrule
        \bottomrule
    \end{tabular}
\end{table}
\begin{table}[H]
    \centering
    \caption{Maximum and time-averaged relative error of the magnetic energy in the laminated core (Scenario B2)}
    \label{tab:energy_scenario_B2}
    \begin{tabular}{lcc} 
        \toprule
        \toprule
        $\ff$ (\si{\hertz}) & max. error & av. error \\ 
        \midrule
        50 & $\SI{2.2}{\percent}$ & $\SI{1.0}{\percent}$ \\
        100 & $\SI{2.6}{\percent}$ & $\SI{1.3}{\percent}$  \\
        500 & $\SI{3.5}{\percent}$ & $\SI{1.9}{\percent}$ \\
        1k & $\SI{4.7}{\percent}$ & $\SI{2.8}{\percent}$\\
        5k & $\SI{13.9}{\percent}$ & $\SI{4.2}{\percent}$ \\
        10k & $\SI{12.3}{\percent}$ & $\SI{6.4}{\percent}$\\
        \bottomrule
        \bottomrule
    \end{tabular}
\end{table}


\section*{Acknowledgment}
We acknowledge the support from Deutsches Elektronen-Synchrotron DESY.
\ifCLASSOPTIONcaptionsoff
  \newpage
\fi



%
\bibliographystyle{IEEEtran}  
\bibliography{dc_biased_homhbfem}        

%








\end{document}